\documentclass[twocolumn]{aastex631}

\usepackage{booktabs}
\usepackage{amsmath}

\shorttitle{Convection Modeling of $\rm H_2$-Dominated Atmospheres}
\shortauthors{Habib \& Pierrehumbert}

\begin{document}

\title{3D Modeling of Moist Convective Inhibition in Idealized Sub-Neptune Atmospheres}

\correspondingauthor{Namrah Habib}
\email{namrah.habib@physics.ox.ac.uk}

\author[0000-0001-7131-251X]{Namrah Habib}
\affiliation{Atmospheric, Oceanic and Planetary Physics, Department of Physics, University of Oxford, Parks Road, OX1 3PU, Oxford, UK}

\author[0000-0002-5887-1197]{Raymond T. Pierrehumbert}
\affiliation{Atmospheric, Oceanic and Planetary Physics, Department of Physics, University of Oxford, Parks Road, OX1 3PU, Oxford, UK}

\begin{abstract}

Atmospheric convection behaves differently in hydrogen-rich atmospheres compared to higher mean molecular weight atmospheres due to compositional gradients of tracers. Previous 1D studies predict that when a condensible tracer exceeds a critical mixing ratio in H$_2$-rich atmospheres, convection is inhibited leading to the formation of radiative layers where the temperature decreases faster with height than in convective profiles. We use 3D convection-resolving simulations to test whether convection is inhibited in H$_2$-rich atmospheres when the tracer mixing ratio exceeds the critical threshold, while including processes neglected in 1D, e.g. turbulent mixing and evaporation. We run two sets of simulations. First, we perform simulations initialized on saturated isothermal states and find that compositional gradients can destabilize isothermal atmospheres. Second, we perform simulations initialized on adiabatic profiles which show distinct, stable inhibition layers form when the condensable tracer exceeds the critical threshold. Within the inhibition layer, only a small amount of energy is carried by latent heat flux, and turbulent mixing transports a small amount of tracer upwards, but both are generally too weak to sustain substantial tracer or heat transport. The thermal profile gradually relaxes to a steep radiative state, but radiative relaxation timescales are long. Our results suggest stable layers driven by condensation-induced convective inhibition form in H$_2$-rich atmospheres, including those of sub-Neptune exoplanets.

\end{abstract}


%*******************************************************************************
\section{Introduction} \label{sec:intro}

Sub-Neptune exoplanets, planets with radii between $\sim$ 1.5 - 4 Earth radii, are one of the most commonly detected type of exoplanets. However, the nature and interior structure of these exoplanets is still unknown. A commonly accepted model for sub-Neptune structure is a hydrogen-helium-dominated atmosphere over a rocky or icy core \citep{van_eylen_asteroseismic_2018}. However, other possible structure have been proposed and include; ``water worlds" \citep{luque_density_2022} which have high volatile fraction envelops, hot supercritical atmospheres containing a mixture of H$_2$ and H$_2$O \citep{pierrehumbert_runaway_2023, benneke_jwst_2024}, and ``Hycean Worlds" \citep{madhusudhan_habitability_2021}, planets with shallow hydrogen-helium-dominated atmospheres overlaying a water rich interior and large oceans. Understanding the atmospheric state and dynamics of hydrogen-rich atmospheres is crucial to our understanding of sub-Neptune exoplanets. In this work, we explore vertical mixing and the temperature structure of H$_2$-rich atmospheres to inform our understanding of sub-Neptunes and other planets with H$_2$-dominated atmospheres.

In $\rm H_2$-dominated atmospheres, vertical mixing due to convection behaves differently compared to higher mean molecular weight atmospheres, like that of Earth. Any atmospheric tracer will have a higher mean molecular weight than the background atmosphere in H$_2$-rich atmospheres. Therefore, a parcel of pure H$_2$ will be less dense and more buoyant than a parcel containing a mixture of H$_2$ with any atmospheric tracer at the same temperature and pressure. Tracers in H$_2$-dominated atmospheres can strongly stabilize \citep{habib_modeling_2024} and potentially inhibit convection \citep{guillot_condensation_1995, leconte_condensation-inhibited_2017, friedson_inhibition_2017}. \cite{habib_modeling_2024} demonstrated through 3D convection-resolving simulations that compositional gradients of non-condensing tracers (i.e. dry convection) significantly stabilize and affect the atmospheric temperature state in H$_2$-dominated atmospheres.

In moist convection, the atmospheric tracer is near saturation and condenses within the atmospheric temperature–pressure range, releasing latent heat that further drives vertical mixing and impacts the temperature profile. The effects of compositional gradients on moist convection (i.e. moist compositional convection) in H$_2$-dominated atmospheres has primarily been explored using 1D analytical studies \citep{guillot_condensation_1995, li_moist_2015, leconte_condensation-inhibited_2017, friedson_inhibition_2017}. Notably, \cite{guillot_condensation_1995} found that vertical mixing due to moist convection is suppressed when the abundance of an atmospheric tracer exceeds a critical threshold in saturated, H$_2$-dominated atmospheres. 

Following standard meteorological definition, convective inhibition or shutdown refers to the suppression of buoyancy, thus preventing vertical mixing and the initiation of convection \citep{emanuel_atmospheric_1994}. \cite{guillot_condensation_1995} argued that in H$_2$ atmospheres, when convection is inhibited by the gradient of a condensing tracer, heat can only be transferred by radiation or conduction. \cite{leconte_condensation-inhibited_2017, friedson_inhibition_2017} extended the theory of condensation-driven convective inhibition to hypothesize that compositional gradients in H$_2$-rich atmospheres could stabilize the atmosphere against all vertical mixing in the 1D regime, including double diffusive convection. 

\cite{leconte_condensation-inhibited_2017, friedson_inhibition_2017} proposed radiation as the primary mechanism to transport heat though condensation-driven stable regions within the atmosphere. Furthermore, \cite{leconte_condensation-inhibited_2017} argued that in regions where the abundance of a saturated tracer exceeds the critical threshold defined by \cite{guillot_condensation_1995}, a radiative layer forms where the temperature profile is superadiabatic. In superadiabatic regions, temperature decreases with height faster than it would in a convective layer. Radiative layers imply that the deep internal temperature in H$_2$-dominated atmospheres is much hotter than it would be under a purely convective atmospheric profile, assuming the temperature at the top of the atmosphere is the same. 

\cite{leconte_condensation-inhibited_2017} suggested the structure of H$_2$-rich atmospheres consists of a deep dry convective region, overlaid by a radiative superadiabatic region, and topped by a condensing, moist convective region. Several recent studies have applied the hypothesis that superadiabatic layers form due to condensation-driven convective inhibition in H$_2$-rich atmospheres, focusing on Hycean worlds \citep{innes_runaway_2023} and magma ocean planets \citep{misener_importance_2022, markham_convective_2022}. However, determining whether condensation-driven convective inhibition occurs requires 3D models, as 1D theory relies on assumptions that may not be valid in more realistic atmospheric settings.

In this work, we test the hypotheses proposed by \cite{guillot_condensation_1995}, which suggests that convection is suppressed when a saturated condensing tracer exceeds a critical threshold. 1D radiative-convective models allow for fast exploration of planetary structure but often make assumptions that may not be justified in realistic settings. In particular, 1D radiative-convective models typically neglect sub-saturation of the condensable tracer, evaporation of condensates, and turbulent mixing. We aim to explore the robustness of condensation-driven convective inhibition in H$_2$-rich atmospheres while accounting for the processes that are neglected in 1D radiative-convective models. 

In this work, we adopt the standard meteorological definition of convective inhibition: the suppression of buoyancy and consequently convective-driven vertical mixing \citep{emanuel_atmospheric_1994}. Using 3D convection-permitting simulations with Cloud Model 1 v20.3 \cite[CM1][]{bryan_governing_2009}, we explore the vertical temperature and compositional structure in H$_2$-dominated atmospheres when a saturated water vapor tracer exceeds the critical threshold defined by \cite{guillot_condensation_1995}. Our objectives are to answer the following questions:
\begin{itemize}
    \item Is convection inhibited when the tracer abundance surpasses the critical threshold defined by \cite{guillot_condensation_1995}?
    
    \item If convective inhibition regions form, is the temperature profile in these regions superadiabatic? 
    
    \item How are heat and moisture transport impacted if convection is inhibited? 
\end{itemize}
The aim of this work is to understand how condensation-driven convective inhibition is affected by processes neglected in 1D models including turbulent mixing, evaporation of condensates, and sub-saturation. We focus on the suppression and stability of vertical mixing due to convection as a local effect, and leave studying compositional suppression of double diffusive convection \citep[e.g.][]{leconte_condensation-inhibited_2017, friedson_inhibition_2017} and global effects of condensation-driven convective inhibition to future work \citep[e.g.][]{guillot_giant_2022}. 

Recently, \cite{leconte_3d_2024} used 3D convection-permitting simulations to explore the formation of convectively stable regions when tracer abundance exceeds a critical threshold for the sub-Neptune K2~18~b. Their results suggest the development of convectively stable regions, and reproduced the atmospheric structure proposed by \cite{leconte_condensation-inhibited_2017}, where a moist convective layer overlies a radiative superadiabatic region, on top of a dry convection layer. However, \cite{leconte_3d_2024} also found the condensable tracer can be transported through convectively stable regions via turbulent mixing, highlighting the importance of 3D studies. We compare our simulation results with those presented by \cite{leconte_3d_2024} in Section~\ref{sec:cm1_results}. 

In section~\ref{sec:background} we review the condensing compositional convection stability criterion. In section~\ref{sec:cm1_description} we describe our CM1 simulation setup, and present our simulation results and discussions in section~\ref{sec:cm1_results}.

%*******************************************************************************
\section{Theoretical Background} \label{sec:background}

Atmospheric convection is defined as the vertical mixing due to a steady gravitational field acting upon vertical variations in density (i.e. buoyancy) \citep{emanuel_atmospheric_1994}. Typically, in an atmosphere, perturbations in density are either due to temperature or compositional variations. Convective inhibition is defined as the suppression of positive buoyancy and vertical mixing \citep{emanuel_atmospheric_1994}. In this work, we test whether condensing tracers can inhibit convection in H$_2$-rich atmospheres.

The effects of atmospheric tracers on convection (also called virtual effects) are typically represented using virtual temperature, 
\begin{equation}
	T_v = T \left( \frac{1+ q_v/\epsilon}{1+ q_v} \right),
\end{equation}
where $T$ is the atmospheric temperature (K), $q_v$ is the vapor mixing ratio of the atmospheric tracer (kg/kg), and $\epsilon = M_v/M_b$ is the ratio of the mean molecular weight of the tracer to the background atmosphere \citep{emanuel_atmospheric_1994}. The subscript $b$, refers to the background atmosphere bulk composition gas, while the tracer, is denoted by subscript $v$ (for ``vapor"). Effectively, $T_v$ represents density in units of temperature and allows to write the ideal gas law as $p = \rho R_b T_v$ where $p$ is the total atmospheric pressure, $R_b$ is the gas constant for the background atmosphere and $\rho$ is the total atmospheric density. The vapor mixing ratio, $q_v$, quantifies how much atmospheric tracer is present relative to the background air and is defined as the ratio of the density of the tracer ($\rho_v$) relative to the background atmosphere density ($\rho_b$), 
\begin{equation}
	q_v = \rho_v/\rho_b.
\end{equation} 
Compositional effects on convection are determined by how much tracer is present and $\epsilon$. If the atmospheric tracer has a lower mean molecular weight than the bulk atmosphere then $\epsilon < 1$ and $T_v > T$, and the atmospheric tracer can \textit{enhance} buoyancy relative to a parcel of pure background atmosphere at the same temperature. Conversely, when the atmospheric tracer is heavier than the background atmosphere $\epsilon > 1$ and the presence of tracers can \textit{suppress} convection. For H$_2$-rich atmospheres with a water vapor tracer $\epsilon = 8.9$, while for water vapor in Earth's atmosphere $\epsilon = 0.6$.

When the tracer does not condense, stability against convection can be approximated by,
\begin{equation}\label{eq:ledoux_criterion}
     \frac{d\ln T_{v, \rm amb}}{d\ln p} \le \beta_{\rm mix},
\end{equation} 
where $T_{v, \rm amb}$ is the virtual temperature profile of the ambient atmosphere, $p$ is the atmospheric pressure, and $\beta_{\rm mix} \equiv \frac{R_{\rm mix}}{c_{p \rm , \; mix}}$ where $R_{\rm mix}$ and $c_{p \rm , \; mix}$ are a representative specific gas constant and heat capacity at constant pressure for the mixture of tracer with the background atmosphere. Eqn.~\ref{eq:ledoux_criterion} is a form of the Ledoux criterion for local stability \citep{ledoux_stellar_1947} and describes the stability against infinitesimal displacements of a parcel containing a mixture of the bulk atmosphere and a non-condensing tracer \citep[e.g.][]{habib_modeling_2024}. 

To derive the condensing convective stability criterion, we must quantify how condensation affects the composition profiles of both the background atmosphere and the rising air parcel. When the atmospheric tracer condenses, a key difference from the non-condensing case is that the vapor mixing ratio is constrained by temperature via the saturation vapor pressure. Assuming the condensed liquid phase is denser than the vapor phase, the Clausius–Clapeyron relation gives the saturation vapor pressure as, 
\begin{equation} \label{eq:clausius_clapeyeron}
    \frac{d \ln e_{\rm sat}}{d \ln T} = \frac{L_v}{R_v T},
\end{equation} 
where $e_{\rm sat}$ is the saturation vapor pressure of the condensable tracer at a given temperature $T$, $L_v$ is the latent heat of vaporization, and $R_v$ is the gas constant of the condensable tracer. For a given temperature, the saturated vapor mixing ratio is 
\begin{equation} \label{eq:sat_mixing_ratio}
    q_{v, \rm sat} = \epsilon \left( \frac{e_{\rm sat}}{p - e_{\rm sat}} \right).
\end{equation}
Saturation at a given $(p, T)$ is determined using the relative humidity, $RH$, which compares the partial pressure of the tracer to its saturation vapor pressure,
\begin{equation}
    RH = \frac{p_v}{e_{\rm sat}} = \left( \frac{q_v}{\epsilon + q_v} \right) \left( \frac{p}{e_{\rm sat}} \right),
\end{equation}
where $p_v$ is the partial pressure of the tracer at $(p, T)$.

Assuming a saturated air parcel is displaced approximately adiabatically without any mixing with its surroundings, the temperature profile of the rising parcel is given by, 
\begin{equation} \label{eq:moist_adiabat}
     \frac{d \ln{T^*_{\rm ad}}}{d \ln{p}} = \beta_{\rm mix} \left[ \frac{1 + q_{v, \rm sat} \left( \frac{L_v}{R_{b} T} \right)} { 1 + q_{v, \rm sat} \left( \frac{L_v}{c_{p, \rm mix} T} \right) \left( \frac{L_v}{R_v T} \right) \left( \frac{1+q_{v, \rm sat}/\epsilon}{1+q_{v, \rm sat}}\right)} \right].
\end{equation}
Eqn.~\ref{eq:moist_adiabat} is the moist adiabat and gives the temperature-pressure profile of a saturated air parcel as it is displaced \citep{emanuel_atmospheric_1994}. Eqn.~\ref{eq:moist_adiabat} assumes that the volume of the condensed phase is negligible compared to the vapor phase of the atmospheric tracer, and all condensed phases are instantaneously removed. We do not consider the compositional gradient of condensed phases in determining atmospheric stability in this work. 

To determine convective stability considering compositional gradients of a condensing tracer, we need to assess how condensation affects the mean molecular weight gradient of both the displaced air parcel and the background atmosphere. As a saturated air parcel is displaced, condensation leads to a change in the composition of the parcel. Additionally, the background saturated atmosphere exhibits a compositional gradient because the vapor mixing ratio changes with altitude, following the saturation vapor pressure curve. The compositional gradient is given by, 
\begin{equation} \label{eq:compositional_gradient}
    \frac{d \ln{\mu}}{d \ln{p}} = q_{v, \rm sat} \left( \frac{1 - 1/\epsilon}{1+q_{v, \rm sat}} \right) \left[ \left( \frac{L_v}{R_v T} \right) \left( \frac{d \ln{T}}{d \ln{p}} \right) - 1 \right],
\end{equation}
where $\mu$ is the total mean molecular weight of the atmosphere \citep{guillot_condensation_1995, leconte_condensation-inhibited_2017}. 

Stability to convection can be determined by comparing the density, or virtual temperature, of a displaced saturated air parcel to that of the surrounding saturated atmosphere. To compare the virtual temperature directly, we would need to solve Eqn.~\ref{eq:compositional_gradient} and Eqn.~\ref{eq:moist_adiabat} and account for the change in the composition of the displaced air parcel due to condensation. Alternatively, we consider stability against infinitesimal displacements and compare the gradient of virtual temperature of a saturated parcel to that of the ambient saturated atmosphere where $\frac{d \ln{T_v}}{d \ln{p}} = \frac{d \ln{T}}{d \ln{p}} - \frac{d \ln{\mu}}{d \ln{p}}$. A saturated parcel is stable against infinitesimal displacements when,
\begin{equation} \label{eq:stability_criterion}
    \left( \frac{d \ln{T}}{d \ln{p}} - \frac{d \ln{\mu}}{d \ln{p}} \right) \Bigg\rvert_{\rm amb} - \left( \frac{d \ln{T}}{d \ln{p}} - \frac{d \ln{\mu}}{d \ln{p}} \right) \Bigg\rvert_{\rm parcel} \le 0
\end{equation}
where $\left( \frac{d \ln{T}}{d \ln{p}} \right) \big\rvert_{\rm parcel}$ = $\frac{d \ln{T^*_{\rm ad}}}{d \ln{p}}$ is the temperature gradient of the saturated parcel given by the moist adiabat (Eqn.~\ref{eq:moist_adiabat}), and $\left( \frac{d \ln{T}}{d \ln{p}} \right) \big\rvert_{\rm amb}$ is the temperature gradient of the background atmosphere. $\frac{d \ln{\mu}}{d \ln{p}}$ is given by Eqn.~\ref{eq:compositional_gradient} and can be evaluated for the parcel and ambient atmosphere using $\frac{d \ln{T^*_{\rm ad}}}{d \ln{p}}$ and $\left( \frac{d \ln{T}}{d \ln{p}} \right) \big\rvert_{\rm amb}$ respectively. 

Simplifying Eqn.~\ref{eq:stability_criterion} using Eqn.~\ref{eq:compositional_gradient} yields, 
\begin{equation} \label{eq:guillot_criterion}
    \left( \nabla_{\rm amb} - \nabla^*_{\rm ad} \right) \left( 1 - \left( \frac{q_{v, \rm sat}}{1 +  q_{v, \rm sat}} \right) (1 -1/\epsilon) \left( \frac{L_v}{R_v T} \right) \right) \le 0, 
\end{equation}
where $\nabla = \frac{d\ln{T_x}}{d\ln{p}}$, ``amb" represents the atmospheric temperature gradient and the ``$^*$ad" represents the temperature gradient defined by the moist adiabat. Eqn.~\ref{eq:guillot_criterion} is the Guillot criterion for local stability against condensing compositional convection \citep{guillot_condensation_1995}. 

When the atmospheric tracer is lighter than the background atmosphere so that $\epsilon~<~1$, the mean molecular weight term, $(1- 1/\epsilon)$, will always be negative meaning that the compositional term in Eqn.~\ref{eq:guillot_criterion} (second bracket) is always positive. Therefore, in the $\epsilon~<~1$ case, convection occurs when $\nabla_{\rm amb} > \nabla^*_{\rm ad}$. 

Conversely, when $\epsilon~>~1$, a compositional gradient can stabilize an unstable atmospheric temperature. Consider $\nabla_{\rm amb} > \nabla^*_{\rm ad}$, in the absence of a compositional gradient the ambient temperature profile is unstable and convection will occur. However, when accounting for the compositional effects, the compositional gradient can stabilize an unstable atmosphere to infinitesimal displacements and inhibit convection when $q_{v, \rm sat} \geq q_{\rm crit}$ where
\begin{equation} \label{eq:critical_mmr}
    q_{\rm crit} = \frac{R_v T}{L_v (1 -1/\epsilon) - R_vT}
\end{equation}
\citep{guillot_condensation_1995}. Compositional gradients of condensable tracers in $\rm H_2$-dominated atmospheres can provide a strong stabilizing mean molecular weight gradient that can inhibit convection when $q_{v} \geq q_{\rm crit}$. 

In this work, we use 3D convection-permitting simulations to explore the atmospheric structure when accounting for compositional gradients of condensing tracers in $\rm H_2$-dominated atmospheres. 3D convection-permitting simulations allow to build upon the theory presented in this section by additionally considering the effects of turbulent mixing, sub-saturation, and evaporation on the atmospheric temperature profile and stable layers. Turbulent mixing is defined as random chaotic turbulent motion that transports heat and mass but is not always representative of the bulk fluid motion. Turbulent mixing differs from convective mixing because convection is buoyancy-driven bulk fluid movement. 

%*******************************************************************************
\section{Methods} \label{sec:cm1_description}

We use Cloud Model 1 \cite[CM1;][]{bryan_governing_2009} to perform idealized simulations of condensing compositional convection. CM1 is a non-hydrostatic convection-resolving model that has been extensively used to study atmospheric flows. In this work, we utilize CM1 version 20.3 from \cite{habib_modeling_2024}, which incorporates modifications to the original code that include setting the thermodynamic constants and planetary parameters using an input file, and revising the buoyancy equation to account for non-dilute concentrations of atmospheric tracers. An overview of the CM1 equations is provided in the Appendix and a detailed description of the model is presented by \cite{bryan_governing_2009} and \cite{bryan_sensitivity_2012}.

\begin{table*}[ht]
    \caption{Atmospheric constants used within the CM1 condensing compositional convection simulations}
    \centering
    \begin{tabular}{llllll}
        \toprule
        {} & {$\rm c_p$} & {$\rm c_v$} & {R} & {$\rm \varepsilon = M_{v}/M_b$} & {$\kappa$} \\
        {} & {J kg$^{-1}$ K$^{-1}$} & {J kg$^{-1}$ K$^{-1}$} & {J kg$^{-1}$ K$^{-1}$} & {} & {$\rm m^2 kg^{-1}$} \\ 
        \midrule
        Hydrogen (Background Atmosphere) & 14304.0 & 10160.0 & 4144.0 & 8.94 & $1.18 \times 10^{-4}$ \\
        Water Vapor (Condensing tracer) & 1870.0 & 1408.5 & 461.5 & N/A & 0.01 \\ 
        \bottomrule
    \end{tabular}
    \label{table:greycond-atmos-params}
\end{table*}

The thermodynamic constants used in this study are listed in Table~\ref{table:greycond-atmos-params}. Gravity is set to 9.8 m/s$^2$ in all simulations. CM1 is run using the fully compressible equations in Large-Eddy Simulation (LES) mode. Sub-grid scale turbulent mixing is included using a turbulent kinetic energy (TKE) closure \citep{deardorff_stratocumulus-capped_1980} to parameterize turbulent mixing at unresolved scales similar to previous studies \citep[e.g.,][]{tan_convection_2021, lefevre_cloud-convection_2022, habib_modeling_2024}. The TKE scheme in CM1 is adapted to the user-imputed thermodynamic constants.

We use a rigid bottom boundary condition with a fixed surface temperature, and horizontal periodic boundary conditions. The turbulent heat and momentum exchange at the bottom boundary is calculated using the original CM1 formulation. Additionally, a Rayleigh damping layer is applied in the top 10 km of the domain to damp out any upward propagating gravity waves. The timestep is set to 3 seconds and radiative calculations are run once per minute. 

We use the \cite{rotunno_airsea_1987} condensation scheme to account for condensation of the water vapor tracer, rainout and evaporation. The Rotunno and Emanuel scheme only considers the vapor and liquid phases where the vapor mixing ratio is, $q_v = \frac{\rho_v}{\rho_b}$, and the liquid condensate mixing ratio is $q_l = \frac{\rho_l}{\rho_b}$. $\rho_l$ is the density of liquid condensate, $\rho_b$ is the background atmosphere density, and $\rho_v$ is the density of the atmospheric tracer in vapor form. In the Rotunno and Emanuel scheme, a distinction is made between \emph{cloud} and \emph{rain} water. Condensate with $q_l \le 1$~g~kg$^{-1}$ is cloud condensate and has a terminal fall velocity of V = 0 $\rm m \; s^{-1}$, while condensates with $q_l > 1$~g~kg$^{-1}$ is rain condensate and has a terminal fall velocity of V = $-$10 $\rm m \; s^{-1}$. 

Condensation is modeled as a two-step process. First, precipitation occurs \emph{in~situ} and second, the condensate is removed by rainout if $V < 0$. Condensation occurs at a given timestep when $q_v(p) > q_{v, \rm sat}(p)$ where $q_{v, \rm sat}(p)$ is the saturation vapor mixing ratio at the given $(T, p)$. Condensation occurs to remove excess water vapor relative to saturation such that after condensation $q_v(p)~=~q_{v, \rm sat}(p)$. When condensation occurs, a small amount of liquid condensate forms, the vapor pressure of the tracer is adjusted to the saturation vapor pressure, and the latent heat release of condensation causes the local temperature to increase slightly. 

Rainout moves condensates to lower atmospheric layers, where they evaporate if the layers are sub-saturated. Condensates evaporate in a grid cell if $q_v(p) < q_{v, \rm sat}(p)$ and $q_l > 0$. Evaporation occurs until the condensate is depleted or the layer reaches saturation, $q_v(p) = q_{v, \rm sat}(p)$, depending on which occurs first. If a layer is saturated or reaches saturation before the condensate is depleted, the condensate that remains is passed to the next lower layer, and the process repeats layer by layer to the surface. When evaporation occurs, the temperature decreases slightly due to the latent heat required to evaporate the liquid, and the vapor pressure of the tracer is again adjusted accordingly. Detailed cloud condensation growth or microphysics are not considered.

The TWOSTR \citep{kylling_reliable_1995} numerical package is used to solve the gray gas radiative transfer equations. TWOSTR is a two stream plane-parallel gray gas radiative transfer model. A constant background opacity and a constant water vapor opacity are set to $\kappa_b = 1.18 \times 10^{-4} \; \rm{m^2 \; kg^{-1}}$  and $\kappa_v = 0.01 \; \rm{m^2 \; kg^{-1}}$ respectively, based on previous studies \citep{menou_atmospheric_2011, freedman_gaseous_2014, innes_runaway_2023}. The total opacity $\kappa$ varies as a function of pressure $p$ and is given by, 
\begin{equation} \label{eq:opacity}
    \kappa(p) = \frac{\kappa_b + \kappa_v * q_v(p)}{1 + q_v}, 
\end{equation}
where $q_v$ is the vapor mixing ratio of the atmospheric tracer, $\kappa_b$ and $\kappa_v$ are the opacity of the background air and atmospheric tracer, respectively. Shortwave heating of the upper atmosphere and scattering are neglected. Additionally, we do not set a distinct cloud opacity and do not consider cloud radiative feedbacks.

The radiation setup differs from sub-Neptune exoplanets, where the atmosphere is largely heated by in-situ shortwave absorption. We opted for a simplified setup where convection can occur, allowing us to capture the effects of radiation on the temperature profile without the complexity and computational demand of a more realistic radiative transfer scheme. The CM1 setup represents a sub-Neptune with a deep interior containing water vapor without a condensed ocean at the bottom. 

We perform three sets of condensing compositional convection simulations in this work: 
\begin{enumerate}
    \item \textbf{Isothermal Simulations}: CM1 is initialized in a saturated isothermal atmospheric state. Without compositional effects, isothermal atmospheres are stable to convection. Destabilization of isothermal layers is an important novel feature. 

    \item \textbf{Adiabatic Simulations}: CM1 is initialized on adiabatic temperature profiles to explore if layer(s) of convective inhibition form where $q_{v, \rm sat} \geq q_{\rm crit}$ and how this impacts the temperature and compositional profile. 

    \item \textbf{Superadiabatic Simulations}: CM1 is initialized on a saturated superadiabatic temperature profile to determine if saturated superadiabatic states are stable over time when $q_{v, \rm sat} \geq q_{\rm crit}$.
\end{enumerate}
The CM1 setup for the isothermal, adiabatic, and superadiabatic simulations are summarized in Table~\ref{table:greycond-simulation-params}.

\begin{table*}[h!t]
    \caption{Initial atmospheric sounding parameters and domain size used for the CM1 condensing compositional convection simulations. Note, $L_x$ and $L_z$ are the horizontal domain width and vertical domain height respectively. The horizontal domain is a square such that  $L_x$ = $L_y$, and the horizontal resolution is the same in the x and y direction. Simulations are listed with alphabetical identifier that corresponds to later plots. Note isothermal simulations and adiabatic simulations are plotted separately.}
    \label{table:greycond-simulation-params}
    \centering
    \begin{tabular}{@{}llccccccccc@{}}
    \toprule
        & & $\rm p_s$ & $\rm T_s$ & $\rm q_{v,s}$ & $\rm q_{crit}$ & $\rm L_x$ & $\rm L_z$ & nx & nz & Time \\
        & & [bar] & [K] & [kg/kg] & [kg/kg] & [km]  & [km] & & & [days] \\ 
    \midrule
        \multicolumn{10}{l}{\textbf{Initial Isothermal Simulations }}   \\ 
    \midrule
        (a) & No Radiation and $q_v \approx q_{\rm crit}$ & 1  & 273 & 0.055 & 0.057 & 1152 & 350 & 192 & 70  & 100   \\
        (b) & $q_v \approx q_{\rm crit}$ & 1  & 273 & 0.055 & 0.057 & 1152 & 350 & 192 & 70  & 100   \\
        (c) & $q_v > q_{\rm crit}$       & 1 & 300 & 0.321 & 0.062 & 1152 & 150&  192 & 30  & 100  \\
        (d) & $q_v < q_{\rm crit}$       & 20 & 300 & 0.016 & 0.062 & 1152 & 520 & 192 & 104 & 100  \\
    \midrule
        \multicolumn{10}{l}{\textbf{Initial Adiabatic Simulations}}   \\ 
    \midrule
        (a) & Dry then Moist Adiabat & 1 & 450 & 0.2 & 0.1 & 1200 & 500 & 192 & 250 & 1750  \\
        (b) & Moist Adiabat          & 1 & 300 & 0.2 & 0.1 & 1200 & 400 & 192 & 200 & 600  \\
        (c) & Superadiabatic        & 1 & 315 & 0.8 & 0.1 & 384 & 200 & 192 & 200 & 600  \\
    \bottomrule
    \end{tabular}
    \smallskip 
    \raggedright
\end{table*}

\subsection{Isothermal Simulations}

We performed a set of four isothermal simulations. All simulations were initialized with the assumption that the atmosphere is saturated throughout. Given that the initial temperature is constant across the domain, a saturated isothermal profile implies an increase in the vapor mixing ratio as pressure decreases. This effect is mathematically represented in Eqn.~\ref{eq:sat_mixing_ratio}. Here, if $T$ is constant, then $e_{\rm sat}$ remains constant, and as pressure decreases, $q_{v, \rm sat}$ increases. However, increasing $q_{v, \rm sat}$ indefinitely eventually causes CM1 to become numerically unstable, as it introduces more dense fluid above lighter fluid, creating a strongly unstable initial condition. Having a large amount of dense fluid at the top of the domain leads to rapid initial convective mixing with large velocities that cause the model to go unstable. Therefore, we set an arbitrary cutoff for the top of the atmosphere in these simulations to be at the point where $q_{v, \rm sat} = 1$, corresponding to 50\% water by mass. This definition means that the simulation domain height depends on the isothermal temperature we set for the atmosphere.

For all of our saturated isothermal cases, the initial atmospheric state is unstable to non-condensing compositional convection by the Ledoux criterion. This is because $q_{v}(p)$ is increasing with height, the temperature is constant, and the tracer is denser than the background air, therefore a denser fluid is resting above a lighter fluid. Further, atmospheric stability against moist compositional convection can be determined by considering the Guillot criterion given by Eqn.~\ref{eq:guillot_criterion}. In the case of an isothermal atmosphere, $(\nabla_{\rm amb} - \nabla^*_{\rm ad}) < 0$. Therefore, the initial atmospheric state is locally stable against convection by the Guillot criterion when $q_v(p) \le q_{\rm crit}$ and unstable when $q_v(p) > q_{\rm crit}$.

We perform three saturated isothermal test cases where we set the atmospheric temperature such that the surface saturated vapor mixing ratio, $q_{v, \rm sat}(p_s)$, is approximately equal, greater than, and less than the surface critical vapor mixing ratio $q_{\rm crit}(T_s)$ given by Eqn.~\ref{eq:critical_mmr}. Specifically, we run simulations at 
\begin{enumerate}
    \item $T$~=~273 K, $p_s$~=~1~bar, and $q_{v, \rm sat}(p_s)~\approx~q_{\rm crit}(T_s)$.
    
    \item $T$~=~300~K, $p_s$~=~1~bar, and $q_{v, \rm sat}(p_s)~>~q_{\rm crit}(T_s)$,
    
    \item $T$~=~300~K, $p_s$~=~20~bar, and $q_{v, \rm sat}(p_s)~<~q_{\rm crit}(T_s)$,
\end{enumerate} 
We also run the 273 K test case without radiation as a control test case where the final state is solely representative of condensing compositional convection. For the case where $q_{v, \rm sat}(p_s) < q_{\rm crit}(T_s)$, the surface pressure was adjusted to maintain the atmospheric temperature in a regime where condensation occurs while ensuring that the initial condition $q_{v, \rm sat}(p_s) < q_{\rm crit}(T_s)$ was met.

In the $q_{v, \rm sat}(p_s) > q_{\rm crit}$ simulation, the initial atmospheric state is unstable to both moist and dry compositional convection. Conversely, in the $q_{v, \rm sat}(p_s) < q_{\rm crit}$ the initial atmospheric state is locally stable to condensing compositional convection by the Guillot criterion in the region where $q_{v, \rm sat}(p_s) < q_{\rm crit}$ and unstable to dry compositional convection by the Ledoux criterion. In the $q_{v, \rm sat}(p_s) < q_{\rm crit}$ case we are testing if condensation can stabilize the Ledoux unstable atmospheric state. The initial conditions were chosen to highlight the compositional destabilization and the nature of evolution towards an end state. The domain size, resolution, and simulation time for the CM1 test cases are given in Table~\ref{table:greycond-simulation-params}. 

Recent studies report deep, radiative isothermal layers in magma-ocean planets \citep[e.g.,][]{nicholls_convective_2025} and in pure-steam atmospheres \citep[e.g.,][]{selsis_cool_2023}. Here we examine the stability of isothermal layers to compositional gradients. If outgassing from a liquid-water or magma ocean allows for the buildup of condensable in the deep atmosphere, can the resulting mean molecular weight stratification maintain these isothermal layers over time? In the absence of compositional effects, an isothermal profile is statically stable to convection ($(\nabla_{\rm isothermal} - \nabla^*_{\rm ad}) < 0$). Using CM1, we explore how compositional gradients affect isothermal layers. We expect compositional gradients to destabilize isothermal layers due to the Ledoux criterion. However, in the $q_{v, \rm sat}(p_s) < q_{\rm crit}$, the initial state is stable to condensing compositional convection. The the $q_{v, \rm sat}(p_s) < q_{\rm crit}$ simulation will test whether condensation can stabilize the Ledoux unstable stratification and preserve deep isothermal layers.

\subsection{Adiabatic Simulations}

We performed two adiabatic condensing compositional convection simulations, which were initialized: (1) with the entire atmosphere set to a moist adiabat defined by Eqn~\ref{eq:moist_adiabat}, and (2) with the lower atmosphere following a dry adiabat with a constant $q_v = 0.2$ kg/kg until the point of saturation, after which the temperature profile transitions to a moist adiabat. When the temperature in the upper atmosphere reached 150 K, the temperature profile was set to be isothermal emulating a radiative stratosphere. The dry adiabat describes the temperature profile of an atmosphere that is well-mixed with a tracer that is either sub-saturated or non-condensing in the temperature-pressure range of the atmosphere. The dry adiabat is defined as, 
\begin{equation} \label{eq:dry_adiabat}
    \frac{d\ln{T_{\rm ad}}}{d\ln{p}} = -\frac{R_{\rm mix}}{c_{p, \rm \; mix}},
\end{equation}
where $R_{\rm mix}$ and $c_{p \rm , \; mix}$ are the specific gas constant and heat capacity for the well-mixed region. The CM1 test case initialized on a dry then moist adiabat is qualitatively a sub-Neptune model with a sub-saturated non-condensing convective deep atmosphere, which becomes saturated and condensing in the colder upper atmosphere. The moisture reservoir is the sub-saturated deep layer, rather than a liquid ocean. Both adiabatic simulations are initialized such that $q_{v, s} > q_{\rm crit}(T_s)$. 

With the adiabatic simulations, we aim to determine whether a superadiabatic temperature profile will form where $q_{v, s} > q_{\rm crit}(T_s)$ and convection is inhibited by the Guillot criterion. The domain size, resolution, and simulation duration for the CM1 test cases are summarized in Table~\ref{table:greycond-simulation-params}. 

\subsection{Superadiabatic Simulation}

Lastly, we perform a single simulation that is initialized on a saturated superadiabatic temperature profile defined by, 
\begin{equation} \label{eq:superad}
    T = T_s \left(\frac{1 + e^{-z/H}}{2}\right)^{1/4},
\end{equation} 
where $T_s$ is the surface temperature set to 315 K, $z$ is the domain height, and $H$ is a reference vertical scale height that was set to 26 km. The temperature profile was set such that the initial state is superadiabatic, the temperature decreases faster with height compared to an adiabatic profile, in the lower atmosphere. The initial tracer mixing ratio was set such that the the atmosphere is fully saturated in the lower domain. In the upper atmosphere where the temperature profile becomes isothermal, the vapor mixing ratio was set to zero in order to avoid the case where the mixing ratio is increasing with height in the isothermal region. With the superadiabatic simulation, we aim to determine if saturated superadiabatic layers remain stable over time as predicted by the Guillot criterion. In the Appendix, we provide a the results of a CM1 simulation initialized on the same superadiabatic profile given by Eqn.~\ref{eq:superad}, but without any condensation and moisture (dry) to benchmark that the model does evolve to a dry adiabat in the absence of condensation.

%*******************************************************************************
\section{Results \& Discussion} \label{sec:cm1_results}

\subsection{Initially Isothermal Simulations}
\label{sec:isothermal}

\begin{figure*}[h!t]
	\centering 
	\includegraphics[width=\textwidth]{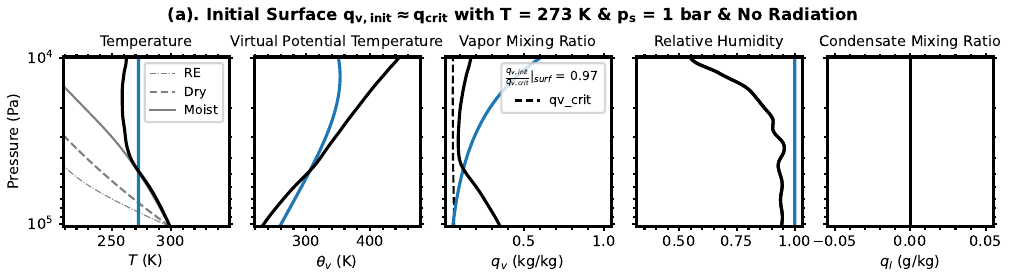}
	\includegraphics[width=\textwidth]{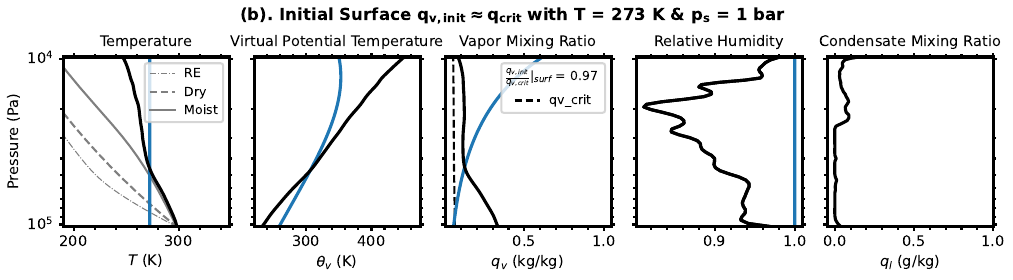}
	\includegraphics[width=\textwidth]{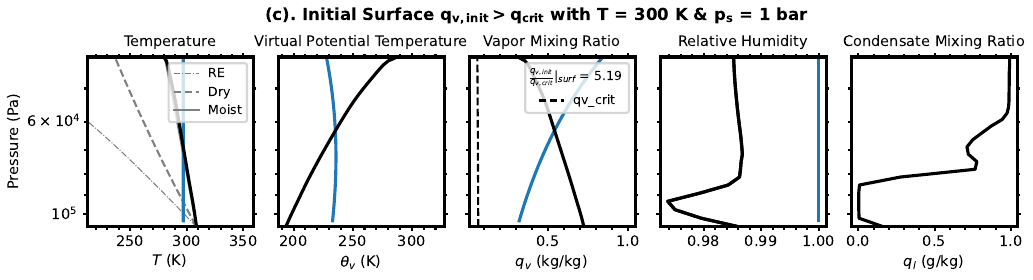}
	\includegraphics[width=\textwidth]{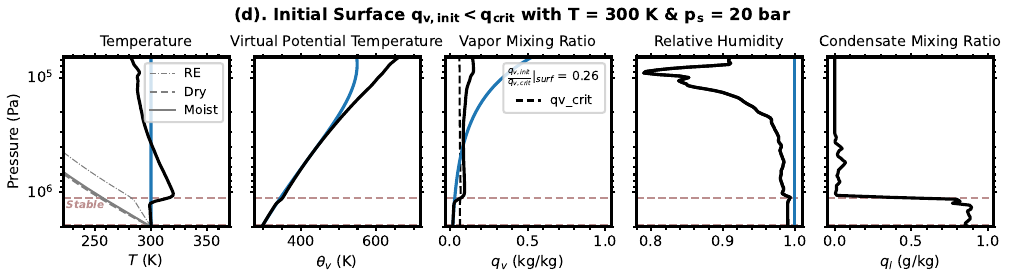}
	\caption{Overview of the atmospheric state from the initially isothermal simulations. In each row, the panels show from left to right, profiles of temperature (K), virtual potential temperature (K), vapor mixing ratio (kg/kg), relative humidity and condensate mixing ratio (g/kg). In the temperature panels, the gray dashed and solid lines plot a dry and moist adiabats, and the gray dot-dashed line plots the temperature profile assuming gray gas pure radiative equilibrium. The dashed gray line in the vapor mixing ratio panels plots the critical vapor mixing ratio of the final state. Blue lines show the initial state while black lines show the final state. In the third row, the final state in the temperature profile lies perfectly on top of the gray moist adiabat curve throughout the domain. The ratio $q_{v,\rm init}/{q_{v,\rm crit}}$ is calculated using the surface values of the initial state of the respective simulation. No initial state is shown in the $q_l$ panels as this was set to be zero everywhere in all the simulations.}
	\label{fig:isothermal-results}
\end{figure*}

\begin{figure*}[h!t]
	\centering 
	\includegraphics[width=1.0\textwidth]{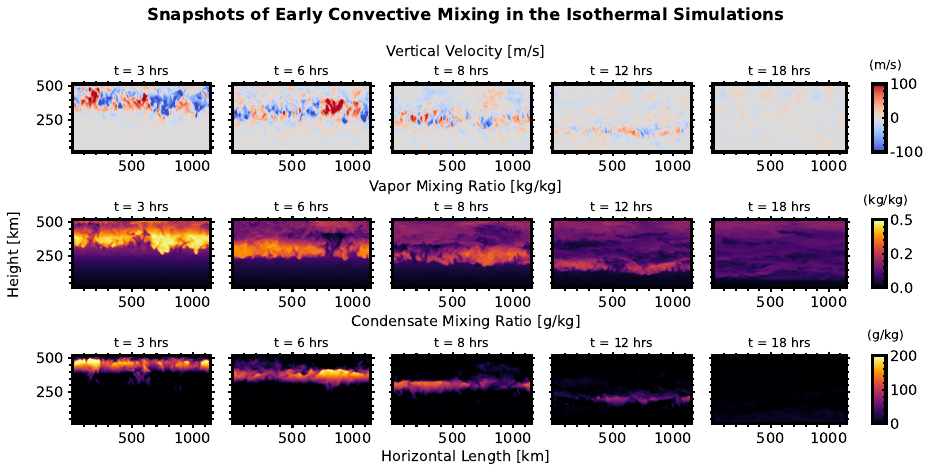}
	\caption{Snapshots of vertical velocity (top row), vapor mixing ratio (middle row), and condensate mixing ratio (bottom row) during the early rapid convective mixing phase in the isothermal simulations. All four isothermal simulations undergo a similar rapid convective mixing phase within the first 12-15 hours. Here, we present the early mixing of the $q_{v,s}~<~q_{\rm crit}$ test case. Mixing initiates near the top of the domain, driven by dense parcels sinking. Within the convecting layer, as compensating rising air parcels become supersaturated, water vapor condenses within them. The condensate then falls through lower cells and evaporates in sub-saturated grid cells. The convecting layer moves down through the atmosphere until it reaches the surface or a neutrally stable layer. In the isothermal simulation initialized with $q_{v,s}~<~q_{\rm crit}$ (shown here), the convecting layer stops and stabilizes at the top of the Guillot stable region rather than reaching the surface as in the other three isothermal simulations. After the initial rapid mixing, the isothermal simulations reach a convectively stable state, which is subsequently altered on a much slower timescale by radiative processes.}
	\label{fig:isothermal-snapshots}
\end{figure*}

Fig.~\ref{fig:isothermal-results} shows the results from the four isothermal CM1 simulations in each respective row. In each row, the figure shows profiles of (from left to right) temperature (K), virtual potential temperature (K), vapor mixing ratio (kg/kg), relative humidity and condensate mixing ratio (g/kg). In the first three panels, the blue line shows the initial state of the simulation and the black line shows the final state. In the temperature panel, the gray dot-dashed line plots a temperature profile calculated in pure radiative equilibrium, the gray dashed line is the dry adiabat defined by Eqn.~\ref{eq:dry_adiabat}, and the gray solid line plots a moist adiabat defined by Eqn.~\ref{eq:moist_adiabat}. Pure radiative equilibrium is calculated assuming a gray gas atmosphere using, 
\begin{equation} \label{eq:temp_rad_equil}
    \frac{d\ln{T_{\rm RE}}}{d\ln{p}} = -\frac{1}{4} \frac{1}{(1 + \tau_\infty - \tau)} p \frac{d\tau}{dp},
\end{equation}
where $T_{\rm RE}(p)$ is the temperature profile in radiative equilibrium, $p$ is the atmospheric pressure, $\tau$ is optical depth, and $\tau_\infty$ is the total optical depth of the atmosphere \citep[see Chap.4,][]{pierrehumbert_principles_2010}. The optical depth is related to the atmospheric opacity, $\kappa$, by ${d\tau}/{dp}~=~-{1}/{(2g)}~\kappa(p)$, where the opacity is determined using Eqn.~\ref{eq:opacity} with the final state CM1 vapor mixing ratio profile. In the vapor mixing ratio panel, the black dashed line plots the critical vapor mixing ratio calculated using Eqn.~\ref{eq:critical_mmr} of the final atmospheric state. The pressure, temperature, and vapor mixing ratio values are calculated by taking the horizontal average of the CM1 data at the given time step. The relative humidity is calculated using the 3D temperature, pressure and vapor mixing ratio outputs from CM1, and the horizontal average of the relative humidity is shown. 

The isothermal CM1 simulations are used to initialize the system without biasing the end-state after convective mixing to be adiabatic. Isothermal profiles are typically stable to convection. However, in our simulation setup, the initial state compositional gradient is increasing with altitude, as seen by the blue line in the $q_v$ panels in Fig.~\ref{fig:isothermal-results}. Therefore, in the upper portion of the domain, the initial state is unstable to non-condensing convection, which is shown by the decreasing $\theta_v$ in the upper portion of the domain, as shown by the blue line the virtual potential temperature panels in Fig.~\ref{fig:isothermal-results}. Decreasing virtual potential temperature indicates convective instability. Initially, convection is driven by the Ledoux criterion, where dense fluid sinks from the top of the domain (see Fig.~\ref{fig:isothermal-snapshots}). However, rising plumes are saturated and the convective stability of rising plumes is determined by the Guillot criterion. In the isothermal simulations there is an asymmetry between the stability of rising and sinking plumes. We use CM1 simulations to understand how compositional gradients influence convective mixing, formation of layers where convection is suppressed, and the ways in which radiation can modify the temperature profile. 

First, consider the isothermal test cases where the initial state $p_s$ = 1 bar, $T(p)$ = 273 K and the surface $q_{v, \rm sat} \approx q_{\rm crit}$. In Fig.~\ref{fig:isothermal-results}, panel (a) shows the atmospheric state without the effects of radiation, while panel (b) shows the atmospheric state with radiation considered. Comparing these two rows, we find that the final vapor mixing ratio state is similar. However, there are notable differences in the upper domain final-state temperature and relative humidity profiles. 

When there are no radiative effects, panel (a) in Fig.~\ref{fig:isothermal-results}, temperature follows a moist adiabat from the surface up until a point in the upper atmosphere where the temperature starts to increase and becomes more isothermal. The region of the atmosphere described by the moist adiabat corresponds to a relative humidity near saturation. In the upper part of the atmosphere, the relative humidity shows the atmosphere is sub-saturated. The temperature profile's shape in the upper atmosphere is similar to the observed shape of the temperature profile of non-condensing compositional convection simulations in \cite{habib_modeling_2024}. 

Convection is initially driven by compositionally dense air sinking from the top of the domain and heating the atmosphere adiabatically until the parcels reach the surface and can no longer sink. Simultaneously, compensating light air parcels rise and cool the atmosphere adiabatically relative to the initial state. Since the initial state was saturated, as the air parcels rise, they become supersaturated, leading to condensation and latent heat release. This process slightly warms the atmosphere, enhances mixing, and explains why the observed cooling in the upper atmosphere is not as strong as in the dry compositional convection results presented by \cite{habib_modeling_2024}. Condensates, formed with $q_l~>~1$~g/kg, fall through the lower atmosphere and evaporate in sub-saturated grid cells, maintaining saturation in the lower atmosphere. Mixing in the lower atmosphere adjusts the temperature profile to a moist adiabat.

The final atmospheric state profiles in panel (a) of Fig.~\ref{fig:isothermal-results} shows the indicative atmospheric state after the initial rapid convective mixing for all four isothermal simulations, as the temperature profile will not continue to change in simulations without radiation. No liquid condensate is observed in the final state in panel (a), suggesting that all condensate formed during the initial rapid mixing phase evaporates before a stable state is reached. In the $q_{v, \rm sat} \approx q_{\rm crit}$ simulation without radiation, we find that the system rapidly adjusted due to convective mixing and a final equilibrium state that is convectively stable is reached. Fig.~\ref{fig:isothermal-snapshots} shows snapshots of the vertical velocity, vapor mixing ratio and condensate mixing ratio taken at five time steps \textit{early-on} within the isothermal simulations to show how mixing occurs during the initial rapid convective adjustment phase. All four isothermal simulations experienced a similar initial rapid adjustment phase that lasted about 12-15 hours. 

In the $q_{v, \rm sat} \approx q_{\rm crit}$ simulation with radiation, shown in panel (b) of Fig.~\ref{fig:isothermal-results}, we find the lower atmospheric temperature profile still follows a moist adiabat. However, the upper atmospheric temperature profile experiences cooling due to radiation, as it relaxes towards a radiative equilibrium stratosphere. Radiative cooling leads to condensate formation in the upper domain after the initial rapid convective mixing phase. Note, the relative humidity in $q_{v, \rm sat} \approx q_{\rm crit}$ case with radiation remains near saturation within the entire column. Radiative cooling continuously produces condensate aloft, which subsequently evaporates and maintains near-saturation in the upper domain, unlike the results in panel (a). The final state for the $q_{v, \rm sat} \approx q_{\rm crit}$ simulation with radiation is stable, with both the compositional and temperature profile stable to convection. 

Panel (c) in Fig.~\ref{fig:isothermal-results} shows the atmospheric state for isothermal test case initialized with $p_s$ = 1 bar and $T(p) = 300$~K, where $q_{v,s} > q_{\rm crit}$. The final atmospheric temperature profile aligns closely with a moist adiabat with the entire domain near saturation. The alignment is so perfect that the moist adiabat curve is not visible in the temperature profile plot in panel (c). Note that for this case, the moist adiabat in a hydrogen atmosphere is itself quite close to being isothermal. The results for $q_{v,s} > q_{\rm crit}$ are similar to those where $q_{v, \rm sat} \approx q_{\rm crit}$ with radiation, but in the $q_{v,s} > q_{\rm crit}$ case, the initial rapid convective mixing is more vigorous, causing the entire domain's temperature profile to evolve to a moist adiabat. As before, after the initial mixing, condensation occurs due to radiative cooling in the upper atmosphere, forming a cloud deck in the upper domain. Overall, the system reaches a convectively stable state where both the temperature and compositional profile are stable to convection. 

In the preceding cases, the Guillot criterion states that superadiabatic profiles, steeper than the moist adiabat, are stable when $q_{v} > q_{\rm crit}$. Given the optical thickness of the atmosphere, radiation would eventually cause the temperature profile to continue to steepen, towards the radiative equilibrium profiles indicated in the figures. However, for optically thick atmospheres the radiative relaxation time is long, and the steepening has not yet had time to take place to any significant extent in these simulations. The simulations provide some insight into the basic nature of radiative-convective dynamics in the compositionally-stabilized regime, with the sequence consisting of a rapid adjustment to the moist adiabat followed by a much slower radiative steepening. 

Panel (d) in Fig.~\ref{fig:isothermal-results} shows the final isothermal CM1 test case initialized with $T = 300$ K $p_s$ = 20 bar, and $q_{v,s}~<~q_{\rm crit}$. In this case, the initial state lower atmosphere, from the surface to $\sim 10^6$~Pa, is stable to moist convection by the Guillot criterion but the entire atmosphere is unstable to dry convection by the Ledoux criterion. Panel (d) in Fig.~\ref{fig:isothermal-results} shows a brown horizontal line dividing the initial Guillot unstable and stable regions within the atmosphere. Interestingly, no convective mixing is observed in the stable region, leaving the final state of this portion of the vertical domain in a saturated isothermal state. The upper atmosphere undergoes an initial rapid mixing phase like the preceding three isothermal simulations. However, in the $q_{v,s}~<~q_{\rm crit}$ case, sinking air parcels stop just below the stable boundary, and no air parcels are observed rising from the stable region. The temperature profile in the upper atmosphere follows a similar trend to the preceding isothermal simulations. 

Unlike the previous three cases, in the $q_{v,s}~<~q_{\rm crit}$ simulation (Fig.~\ref{fig:isothermal-results}(d)), condensing compositional convection is stabilizing the lower domain even in the final atmospheric state. We find that moist compositional stabilization is strong enough to prevent convective mixing in the lower domain. In the lower domain, the compositional profile increases with height and is unstable according to the Ledoux criterion, but is stabilized by condensing compositional convection described by the Guillot criterion. 

Since all three initially isothermal CM1 cases with radiation rapidly mixed to a moist adiabat and thereafter, only evolve slowly by radiative relaxation, we do not run the isothermal simulations further. Instead, in the next section we initialize CM1 on the moist adiabat and run the CM1 simulation longer to better understand the radiative relaxation. 

\subsection{Adiabatic \& Superadiabatic Simulations}
\label{sec:adiabatic}

\begin{figure*}[h!t] 
    \centering 
    \includegraphics[width=\textwidth]{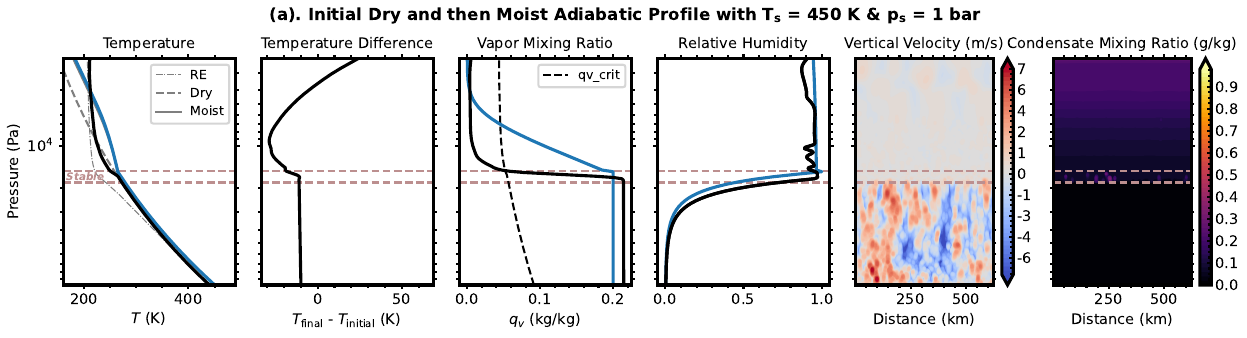}
    \includegraphics[width=\textwidth]{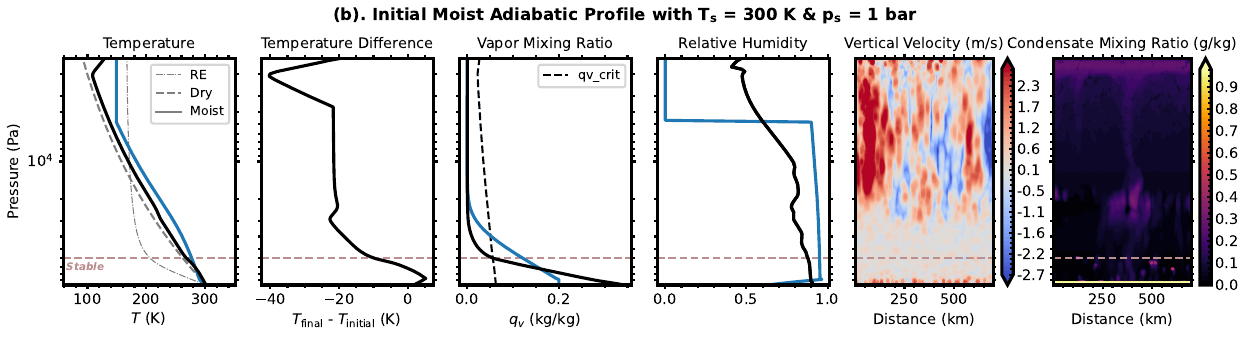}
    \includegraphics[width=\textwidth]{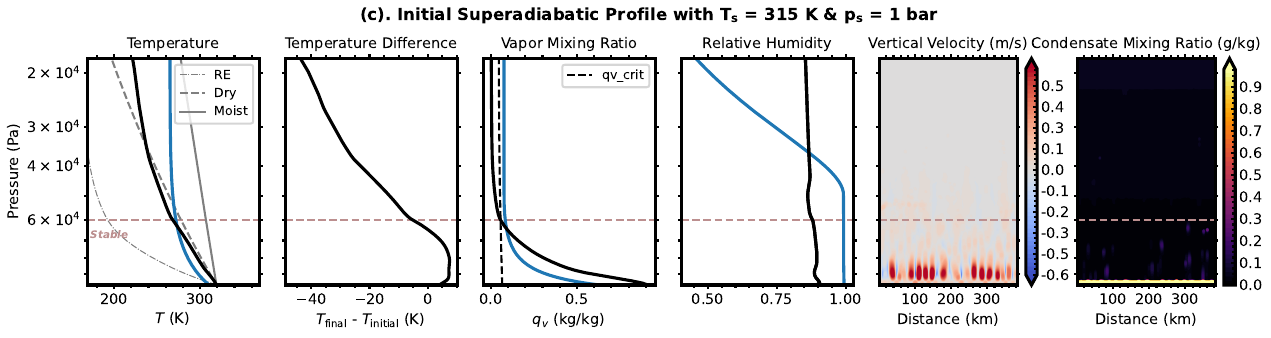}
    \caption{Atmospheric state for the CM1 simulations initialized on (a.) the dry then moist adiabat, (b.) the moist adiabat, and (c.) the superadiabatic state. In each row, the panels show from left to right, profiles of temperature (K), temperature difference between the final and initial state (K), vapor mixing ratio (kg/kg), relative humidity, and vertical cross-sections taken in the middle of the y-domain of vertical velocity (m/s) and condensate mixing ratio (g/kg). In the temperature panels, the gray dashed and solid lines plot a dry and moist adiabats, and the gray dot-dashed line plots the gray gas pure radiative equilibrium temperature profile. The dashed gray line in the vapor mixing ratio panels indicates the critical vapor mixing ratio of the final state. Initial states are illustrated by blue lines in the profile plots, whereas the final state from the CM1 simulation is shown by the solid black line. The cross-sections are shown for the final state. In panels (a) and (b), we observe only the initial stages of steeping towards a pure radiative equilibrium superadiabatic state, as highlighted in the temperature difference profiles. The simulations show long radiative timescales and the results presented here are not in a statistical equilibrium state.}
    \label{fig:adiabatic_results}
\end{figure*}

Fig.~\ref{fig:adiabatic_results} shows an overview of the atmospheric state for the CM1 simulations initialized on the (a.) dry then moist adiabat, (b.) moist adiabat, and (c.) superadiabatic profile. Each row shows panels that display, from left to right, vertical profiles of temperature (K), temperature difference between the final and initial state (K), vapor mixing ratio (kg/kg), relative humidity, and 2D cross-sections taken in the middle of the y-domain of the vertical velocity (m/s) and condensate mixing ratio (g/kg). Blue lines indicates the initial state of the CM1 simulation and the black line depict the final state. The 2D cross-sections are shown for the final state. As before, $T(p)$ and $q_v(p)$, are determined by taking the horizontal average of the 3D CM1 output at the given timestep. Relative humidity is calculated using the 3D CM1 outputs for $T$, $p$ and $q_v$ and the horizontal average is shown. 

Brown dashed horizontal lines in Fig.~\ref{fig:adiabatic_results} are used to highlight boundaries between convective and stable atmospheric layers. The three adiabatic test cases were designed such that moist convection is inhibited according to the Guillot criterion within a region of the atmosphere where $q_v > q_{\rm crit}$. In the simulations initialized on a moist adiabat and superadiabatic state (see Fig.~\ref{fig:adiabatic_results}(b) and (c)), the region stable against moist convection lies below the brown dashed line. In the dry then moist adiabatic simulation, shown by Fig.~\ref{fig:adiabatic_results}(a), the layer where convection is inhibited is located in between the two brown dashed lines. 

The CM1 test case initialized on a moist adiabat represents a continuation of the isothermal cases, which showed rapid mixing that led to the formation of a convectively stable state, characterized by a moist adiabat in the lower domain. The test case starting with a dry then moist adiabat reflects a more realistic scenario where the atmosphere has a deep dry convection zone below an upper moist convecting layer. Lastly, the superadiabatic case with $q_v > q_{\rm crit}$ was performed to test if saturated superadiabatic states remain stable against convection according to the Guillot criterion.

All three simulations presented in Fig.~\ref{fig:adiabatic_results} show condensation-driven convective inhibition. The vertical velocity panel for the moist adiabat simulation, Fig.~\ref{fig:adiabatic_results}(b), clearly shows vertical mixing is near zero from the surface up to pressure of $\sim 2 x 10^4$ Pa. Similarly, we observe near zero vertical velocities in the stable region (between the two brown lines) of the dry then moist adiabatic test case (Fig.~\ref{fig:adiabatic_results}(a)) and within the whole domain of the superadiabatic simulation (Fig.~\ref{fig:adiabatic_results}(c)). The relative humidity stays near saturation in regions that are convectively stable across all three simulations in Fig.~\ref{fig:adiabatic_results}. Further, water vapor is transported from the upper to the lower atmosphere from the initial to final state in all three simulations, as seen by in the increase in final state $q_v$ in the lower portion of the domain. Superadiabatic temperature profiles begin to form in both Fig.~\ref{fig:adiabatic_results}(a) and (b), but neither simulation reaches statistical equilibrium state. Fig.~\ref{fig:adiabatic_results}(c) confirms that saturated superadiabatic states where $q_v > q_{\rm crit}$ remain convectively stable over time according to the Guillot criterion. 

\subsubsection{(a). Dry then Moist Adiabatic CM1 Simulation}

\begin{figure*}
    \centering 
    \includegraphics[width=\textwidth]{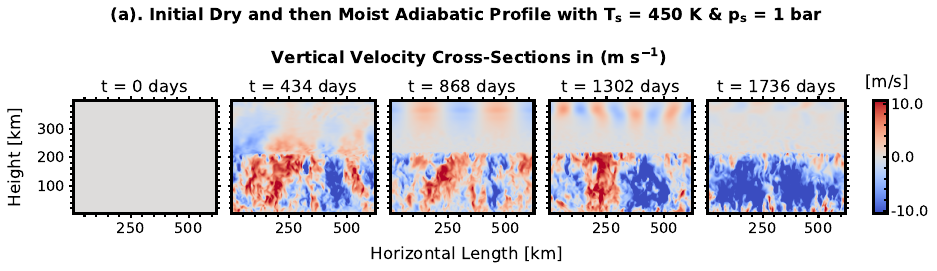}
    \includegraphics[width=\textwidth]{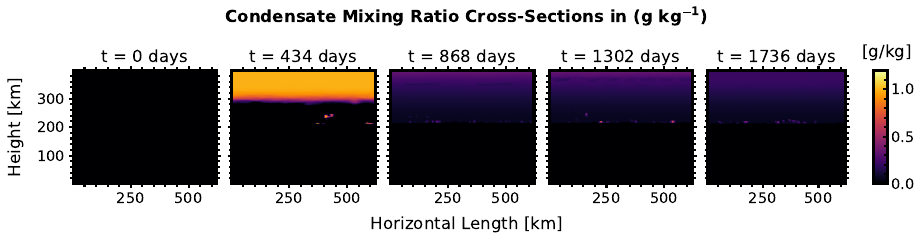}
    \caption{Vertical velocity (top row) and condensate mixing ratio (bottom row) cross-sections in (x,z) taken in the middle of the y-domain for 5 time steps in the CM1 simulation initialized on the dry adiabat to the level of saturation from which point the initial state followed a moist adiabat. The vertical velocity cross-sections show 3 distinct layers; a dry convective region near the surface, a layer where convection is inhibited and velocities are nearly zero, and a layer of weak moist convection in the upper atmosphere. The corresponding condensate mixing ratio plots shows that a cloud deck forms in the moist convection layer and condensates fall through and evaporate in the convective inhibition layer.}
    \label{fig:ad-drythenmoist-velocity}
\end{figure*}

\begin{figure}
    \centering 
    \includegraphics[width=1.0\columnwidth]{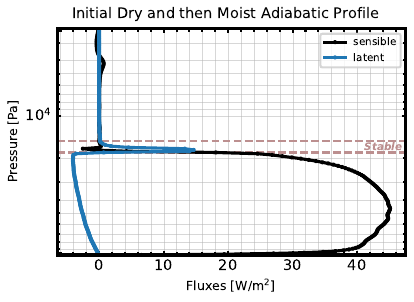}
    \caption{Domain-mean sensible and latent heat fluxes for the CM1 test case initialized on a dry then moist adiabat, averaged over the final 25 days of the simulation. The sharp drop of to zero sensible heat flux within the stable layer indicates suppressed convection. Latent heat flux dominates transport in the condensation-driven convective inhibition layer, and positive $Q_{\rm latent}$ values suggests upward vapor transport driven by evaporation.}
    \label{fig:ad-drythenmoist-fluxes}
\end{figure}

\begin{figure}
    \centering 
    \includegraphics[width=1.0\columnwidth]{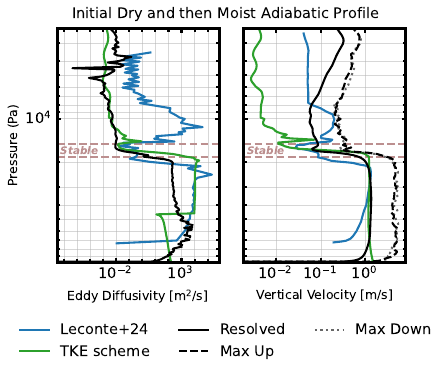}
    \caption{Eddy diffusivity (left panel) and vertical velocities (right panel) of the CM1 simulation initialized on a dry then moist adaibat. In the left panel, the eddy diffusivity from the sub-grid CM1 TKE scheme is shown in green, the resolved $K_{\rm zz}$, calculated using Eqn.~\ref{eq:eddy-diffusivity}, is shown in black, and the eddy diffusivity from \cite{leconte_3d_2024} is shown in blue. In the right panel, the RMS velocity, maximum upward and maximum downward velocity from the CM1 simulation are shown in black solid, dashed and dotted lines respectively. The RMS velocity profile from \cite{leconte_3d_2024} is shown in blue for comparison. As before, the brown dashed lines mark the convectively stable region. Note, the \cite{leconte_3d_2024} profile is shifted upwards such that the convective inhibition layer in their result lies within the brown dashed lines for comparison purposes.}
    \label{fig:ad-drythenmoist-kzz}
\end{figure}

Consider the CM1 simulation initialized on the dry then moist adiabat shown in Fig.~\ref{fig:adiabatic_results}(a). The final state exhibits a three-layer structure: (i) a lower dry convective layer, (ii) a condensation-driven convective inhibition layer where vertical velocity and thus buoyancy and mixing are suppressed but the atmosphere remains near saturation, and (iii) a weak moist convective layer at the top of the domain. The brown dashed lines mark the convective inhibition layer boundaries as predicted by the Guillot criterion. The lower boundary is given by where the atmosphere first reaches saturation and $q_v(p) > q_{\rm crit}$, and the upper boundary is the level where $q_v(p) < q_{\rm crit}$, above which compositional gradients no longer inhibit convection. Interestingly, we observe that vertical velocities and convection remain suppressed above the predicted upper boundary of the convective inhibition layer (top brown dashed line, see also Fig.~\ref{fig:ad-drythenmoist-velocity}). 

Fig.~\ref{fig:adiabatic_results}(a) shows the deep atmosphere approaches radiative equilibrium, as expected in regions with constant opacity and without pressure broadening. The deep atmosphere is located in a dry convection zone where both the vapor mixing ratio and opacity remain unchanged in our model setup. The thermal structure of the lower atmosphere is unchanged from the initial to the final state, and remains on a dry adiabat. A superadiabatic temperature structure begins to form in the stable region in between the two brown lines, as seen in the temperature difference panel in Fig.~\ref{fig:adiabatic_results}(a). In the upper atmosphere, the temperature profile cools towards a radiative equilibrium state leading water vapor to condense and fall through the atmosphere to the boundary of the convective inhibition layer and the dry convection layer, where it evaporates.

The water vapor mixing ratio panel in Fig.~\ref{fig:adiabatic_results}(a) shows moisture has been transported from the upper atmosphere to the dry convection layer between the initial and final state. Further, the final state vapor mixing ratio sharply decreases in the convectively stable region, indicating reduced mixing. The relative humidity panel shows that both the convective inhibition layer and the moist convection layer in the upper atmosphere remain near saturation, suggesting that the convective stability in the stable layer should be governed by the Guillot criterion given by Eqn.~\ref{eq:guillot_criterion}.

Fig.~\ref{fig:ad-drythenmoist-velocity} shows 2D cross-sections taken in the middle of the y-domain of the vertical velocities (m/s, top row) and corresponding condensate mixing ratio (g/kg, bottom row) for five time steps from the dry then moist adiabatic simulation. Fig.~\ref{fig:ad-drythenmoist-velocity} shows a distinct region where vertical mixing is suppressed separating an upper moist convection zone from a lower dry convection zone. As the simulation progresses, vertical velocities in the moist convective zone weaken due to the transport of water vapor to the dry convection zone. In the weak moist convective layer at the top of the domain, the vertical velocities are small but still greater than the convectively inhibited area. 

Fig.~\ref{fig:adiabatic_results}(a) and Fig.~\ref{fig:ad-drythenmoist-velocity} shows a cloud deck, condensate with $q_l(p) < 1$ g/kg, forms early in the simulation and persists within the moist convective layer. Precipitation with  $q_l(p) > 1$ g/kg also develops in the moist convection layer, which then rains-out and evaporates at the boundary between the stable layer and the dry convection zone. Evaporation at this boundary transports moisture to and triggers convective downdrafts in the dry convection layer. The evaporation level defines the lower boundary between the stable and dry convective layers. 

To further diagnose the atmospheric layers, convective mixing, and moisture transport, Fig.~\ref{fig:ad-drythenmoist-fluxes} shows the domain- and time-mean average latent and sensible heat fluxes for the dry then moist adaibat CM1 simulation. The sensible heat flux is defined by 
\begin{equation} \label{eq:sensible_hf}
    Q_{\rm sensible} = c_p \langle \rho w' T' \rangle
\end{equation}
and the latent heat flux is given by, 
\begin{equation} \label{eq:latent_hf}
    Q_{\rm latent} = L_v \langle \rho w' q_v' \rangle
\end{equation}
where $c_p$ is the specific heat capacity, $L_v$ is the latent heat of vaporization, $\rho$ is the total atmospheric density, $w'$ is the vertical velocity perturbation, $T'$ is the temperature perturbation, $q_v'$ is the vapor mixing ratio perturbation, and $\langle \rangle$ denote spatial and temporal averages. Eqns.~\ref{eq:sensible_hf}~--~\ref{eq:latent_hf} are calculated for the last 25 days using the vertical velocity, temperature, and mixing ratio perturbations and density output from CM1. 

In the dry convection layer, the sensible heat flux in Fig.~\ref{fig:ad-drythenmoist-fluxes} is large and positive, indicating strong upward mixing of warm, sub-saturated air in the dry convective layer. Conversely, the latent heat flux is weakly negative ($\sim 0-4$~W~m$^{-2}$), implying a net downward transport of moist air that transports moisture from the upper atmosphere to the dry convection layer, consistent with the $q_v(p)$ profile in Fig.~\ref{fig:adiabatic_results}(a). Additionally, this weakly negative latent heat flux, slightly cools the dry convection layer in agreement with the temperature profile shown in Fig.~\ref{fig:adiabatic_results}(a). 

The stable layer (between the brown dashed lines) in Fig.~\ref{fig:ad-drythenmoist-fluxes} shows a sharp decline to slightly negative values ($\sim$3~W~m$^{-2}$) in the sensible heat flux suggesting suppressed convective mixing with weak net downward sinking of plumes, consistent with the Guillot inhibition criterion (Eqn.~\ref{eq:guillot_criterion}). The latent heat flux in Fig.~\ref{fig:ad-drythenmoist-fluxes} peaks to about 16~W~m$^{-2}$ at the base of the stable layer, indicating an upward moisture flux in the stable layer. 

Above the stable layer (top brown-dashed line), the sensible and latent heat fluxes remain near zero up to about 5$\times 10^3$ Pa where small nonzero flux values suggest intermittent, shallow convective mixing, consistent with the weak moist convective layer seen in the vertical velocity panels in Fig.~\ref{fig:adiabatic_results}(a) and Fig.~\ref{fig:ad-drythenmoist-velocity}. Overall, Fig.~\ref{fig:ad-drythenmoist-fluxes} shows downward moisture transport into the dry convective layer from the upper atmosphere, and strong inhibition of mixing in the stable layer and above to about 5$\times 10^3$ Pa. Within the stable layer, the latent heat flux strongly dominates over the sensible heat flux, suggesting that some energy is carried upward by vapor transport, in agreement with \cite{leconte_3d_2024}. We find a latent heat flux of about 16~W~m$^{-2}$ within the convective inhibition layer, which is comparable to the $\sim$20~W~m$^{-2}$ reported by \cite{leconte_3d_2024}. 

Lastly, to understand the impact of turbulent mixing in the 3D simulations, Fig.~\ref{fig:ad-drythenmoist-kzz} shows the eddy diffusivity flux in left panel and the RMS vertical velocity profile in right panel. The resolved eddy diffusivity, $K_{\rm zz}$ is defined as
\begin{equation} \label{eq:eddy-diffusivity}
    K_{\rm zz}^{\rm resolved} = - \frac{\langle \rho q_v' w' \rangle}{{\Bar{\rho}} \langle \frac{dq_v}{dz} \rangle},
\end{equation}
where $\rho$ is the total atmospheric density, $q_v$ is the vapor mixing ratio, $w$ is the vertical velocity, $z$ is height, $'$ denote perturbations, and $\langle \rangle$ denote spatial and temporal averages. In Fig.~\ref{fig:ad-drythenmoist-kzz} the eddy diffusivity of the dry then moist adiabatic simulation in plotted in black with $K_{\rm zz}$ profile from \cite{leconte_3d_2024} shown in blue for comparison. Additionally, we plot a green line showing the parameterized sub-grid eddy diffusivity output from CM1. CM1 uses the turbulent kinetic energy (TKE) sub-grid turbulence parameterization \citep[adapted from][]{deardorff_stratocumulus-capped_1980} which uses a predictive turbulent kinetic energy to determine an eddy viscosity and diffusivity to represent mixing on sub-grid scales. The TKE scheme has a stability dependence that reduces sub-grid mixing in statically stable conditions. The vertical velocity panel in Fig.~\ref{fig:ad-drythenmoist-kzz} shows the RMS velocity, the maximum upward and downward velocities in black solid, dashed and dotted lines, respectively. The RMS velocity profile from \cite{leconte_3d_2024} is shown in blue for comparison and the sub-grid velocity from the TKE scheme is shown in green. Note that the profiles from \cite{leconte_3d_2024} are shifted so that the stable regions of the two studies align for easier comparison.  

Fig.~\ref{fig:ad-drythenmoist-kzz} supports that convection is inhibited in the stable layer up to about 5$\times 10^3$ Pa. In the stable layer between the brown dashed lines, both the resolved (black line) and sub-grid TKE parameterized (green line) eddy diffusivity and RMS velocities decrease by several orders of magnitude further indicating convective inhibition. Both the RMS velocity and $K_{\rm zz}$ only recovers weakly aloft near the top of the domain, corresponding to the weak moist convective layer. The small resolved (black-line) sub-grid eddy fluxes and RMS velocity suggest that moist convection at the top of the domain is driven by radiative cooling creating condensation. The resolved $K_{\rm zz}$ and RMS velocity do not recover or increase above the stable layer unlike the $K_{\rm zz}$ profile from \cite{leconte_3d_2024} which shows a strong increase in the upper atmosphere corresponding to a dynamically driven moist convective layer. Note, \cite{leconte_3d_2024} did not use or employ a sub-grid scale turbulence model, therefore, the blue profile is comparable to the resolved (black-line) from our study.

Immediately above the stable layer, the TKE sub-grid $K_{\rm zz}$ and RMS velocity do increase suggesting sub-grid, small-scale turbulent mixing above the stable layer that transport water vapor upwards. However, unlike the findings in \cite{leconte_3d_2024}, our results indicate that diffusive fluxes are insufficient to transport enough water vapor upwards from the stable region to sustain strong moist convection. 

Our model shows similar profile shape and magnitude RMS vertical velocity and $K_{\rm zz}$ in the lower atmosphere and convective inhibition layer compared to findings in \cite{leconte_3d_2024}. Both the resolved $K_{\rm zz}$ and the RMS vertical velocity remain comparable to the stable convective inhibition region above the upper boundary of the stable region, suggesting a transition or boundary layer where convection still seems to be suppressed even though $q_v(p) < q_{crit}$.  The upper boundary of the stable region is defined by $q_v(p) = q_{crit}$, beyond which the Guillot criterion no longer predicts convective inhibition. A transition or boundary region was not found in the results presented by \cite{leconte_3d_2024}. 

In general, the $K_{\rm zz}$ and the vertical velocity profiles suggest that convection is still suppressed above the stable layer within the CM1 results, supporting earlier observations that moist convection is weakening in the upper atmosphere and the temperature profile is evolving towards radiative equilibrium. We attribute the difference in mixing magnitude above the stable layer between this work and \cite{leconte_3d_2024} to variations in model setup, particularly in the condensation and turbulence parameterizations. \cite{leconte_3d_2024} used a different condensation scheme and find higher latent heat fluxes than this study, which could enhance mixing in their model's upper atmosphere. Additionally, we employ a sub-grid turbulence parameterization, whereas \cite{leconte_3d_2024} did not. In the TKE sub-grid parameterization used in CM1, mixing is suppressed in statically stable regions, which could further reduce mixing in the upper atmosphere in our CM1 simulation results. Due to the lower eddy diffusivity and vertical velocities in our model results, we do not transport enough water vapor upwards to sustain strong moist convection in the upper atmosphere unlike \cite{leconte_3d_2024}. 

Overall, the dry then moist adiabatic simulation shows that condensation-driven convective inhibition layers do form in hydrogen-rich atmospheres where $q_v(p) > q_{crit}$. Additionally, we find that the inhibition layer extends beyond where $q_v(p) \le q_{crit}$ due to a potential boundary layer and weak upward transport of water vapor. Despite the clear formation of a convective inhibition layer, we do not observe the temperature profile adjust to a steep radiative profile. This is because the simulations were quite computationally expensive to run and the radiative relaxation timescales are long. The dry then moist adiabatic CM1 test case confirms that convection is inhibited, thus, radiative relaxation should eventually proceed to produce steep radiative layers in the inhibited layer. In Section~\ref{sec:radiation}, we discuss and quantify the radiative relaxation timescales for all of the CM1 simulations.

\subsubsection{(b). Moist Adiabatic CM1 Simulation}
\begin{figure}
    \centering 
    \includegraphics[width=1.0\columnwidth]{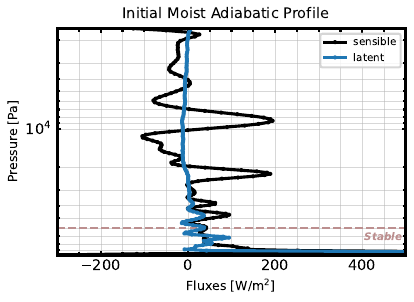}
    \caption{Domain-mean sensible and latent heat fluxes for the CM1 test case initialized on a moist adiabat calculated using Eqns.\ref{eq:sensible_hf}-\ref{eq:latent_hf}. Heat fluxes are averaged over the last 25 days of the simulation.}
    \label{fig:ad-moist-fluxes}
\end{figure}

\begin{figure}
    \centering 
    \includegraphics[width=1.0\columnwidth]{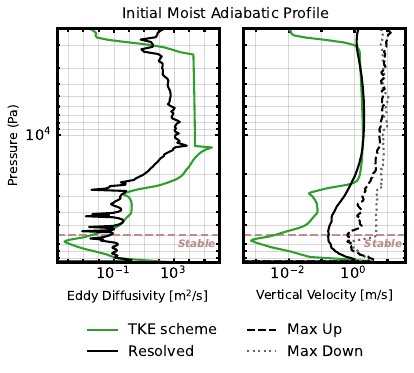}
    \caption{Eddy diffusivity (left panel) and vertical velocities (right panel) of the CM1 simulation initialized on a moist adaibat. In the left panel, the eddy diffusivity from the CM1 test case is calculated using Eqn.~\ref{eq:eddy-diffusivity}. In the right panel, the RMS velocity, maximum upward and maximum downward velocity from the CM1 simulation is shown in black, solid, dashed and dotted lines respectively. The green line plots the eddy diffusivity and vertical velocity output from the CM1 sub-grid turbulence parameterization. As before, the brown dashed lines mark the convectively stable region.}
    \label{fig:ad-moist-kzz}
\end{figure}

Generally, the moist adiabatic (b) and superadiabatic (c) CM1 simulations show trends similar to the dry then moist adiabatic simulation (a). The moist adiabatic simulation, in Fig.~\ref{fig:adiabatic_results}(b) shows clear condensation-driven convective inhibition from the surface up to $\sim$~2~$\times 10^4$~Pa where vertical velocities are near zero and thus buoyancy and convective mixing are suppressed. In the moist adiabatic simulation, the atmosphere has a two layer structure: a stable convective inhibition layer from the surface to pressures nearing $\sim$~2~$\times 10^4$~Pa, topped by a convective layer aloft. As in the previous case, the stable layer seems to extend beyond where the Guillot criterion predicts compositional driven convective inhibition (brown line where $q_v(p) = q_{crit}$), suggesting the presence of a boundary or transition layer. 

Fig.~\ref{fig:adiabatic_results}(b) shows that the temperature profile is cooling towards radiative equilibrium within the whole domain. In the upper atmosphere, the temperature lies between a dry and moist adiabat but appears to be cooling to the radiative equilibrium, while the stable region's thermal profile is steepening over time and also evolving towards a radiative profile, as seen in the temperature difference plot. Moisture is transported from the upper atmosphere towards the surface between the initial and final states. A buildup of moisture at the surface is observed in the condensate mixing ratio cross-section in Fig.~\ref{fig:adiabatic_results}(b). Radiative cooling in the upper atmosphere drives condensation which rains out and evaporates in lower grid cells keeping the upper atmosphere at saturation and driving moist convection in the upper atmosphere. Some condensate is observed to rain out and evaporate in the inhibition layer, however, the transition boundary layer between the moist convective layer and the inhibition layer remains sub-saturated. 

To understand the moisture transport and convective mixing, Fig.~\ref{fig:ad-moist-fluxes} shows the sensible and latent heat fluxes for the moist adiabatic simulation. The latent heat flux peaks at the surface due to enhanced evaporation-driven moisture flux from the buildup of condensates at the surface. Near the top of the stable layer, the latent heat flux peaks again indicating upward moisture flux from the stable layer to the upper atmosphere. Above the inhibition layer, latent fluxes are small and slightly negative suggesting a net downward transport of moist air. Conversely, the sensible heat flux is suppressed in the stable layer with weakly positive values, but shows oscillations in the upper atmosphere consistent with small-scale mixing in the upper atmosphere. The sensible heat flux remains small but does increase above the stable layer supporting the argument for a transition layer above the stable region. 

Fig.~\ref{fig:ad-moist-kzz} shows the $K_{\rm zz}$ (left panel) and RMS vertical velocity (right panel) profiles. The black line shows the resolved $K_{\rm zz}$ calculated using Eqn.~\ref{eq:eddy-diffusivity} in the $K_{\rm zz}$ panel, and RMS vertical velocity in the left panel, while the green line plots the $K_{\rm zz}$ and vertical velocity from CM1's TKE sub-grid turbulence scheme, respectively. Within the stable layer, both the resolved and TKE drop by 3-4 orders of magnitude in $K_{\rm zz}$ (left) and 1-2 orders of magnitude in RMS vertical velocity (right) in Fig.~\ref{fig:ad-moist-kzz}, confirming strong convective inhibition and weak vertical transport across this layer. In the stable layer, the m s$^{-1}$ order of magnitude RMS velocity suggest some convective mixing driven by the condensation and evaporation of condensates in this layer, especially at the surface. The resolved $K_{\rm zz}$ and RMS vertical velocity remains suppressed throughout and beyond the inhibited layer up to about $2 \times 10^4$ Pa, where they then begin to increase consistent with weak mixing and gravity waves. Sub-grid scale mixing from the TKE parameterization increases immediately above the brown line where Guillot criterion predicts convective inhibition boundary, suggesting mixing is not inhibited above the convective inhibition layer but is suppressed. 

As in the previous simulation, we find condensation-driven convective inhibition in the moist adiabatic simulation but do not observe a final state temperature profile on a steep radiative profile, again indicating that radiative processes are slow. Fig.~\ref{fig:ad-moist-fluxes}~--~\ref{fig:ad-moist-kzz} suggest water vapor is transported upward at the top of the convective inhibition layer primarily by sub-grid scale turbulence while moisture is transported from the upper atmosphere to the lower atmosphere by condensates. Over time, the upper atmosphere convection should weaken since both the sub-grid and resolved $K_{\rm zz}$ fluxes are small. The moist adiabatic simulation confirms that convection is inhibited, but radiative relaxation should eventually proceed to produce steep radiative layers. 

\subsubsection{(c) Superadiabatic CM1 Simulation}

\begin{figure}
    \centering 
    \includegraphics[width=1.0\columnwidth]{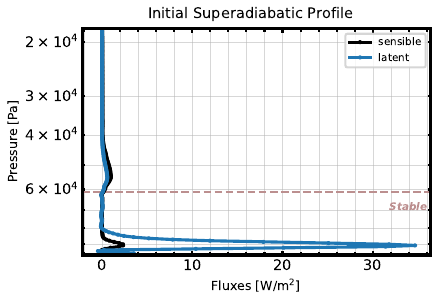}
    \caption{Domain-mean sensible and latent heat fluxes for the CM1 test case initialized on a superadiabatic profile calculated using Eqns.\ref{eq:sensible_hf}-\ref{eq:latent_hf}. Heat fluxes are averaged over the last 25 days of the simulation.}
    \label{fig:ad-superad-fluxes}
\end{figure}

\begin{figure}
    \centering 
    \includegraphics[width=1.0\columnwidth]{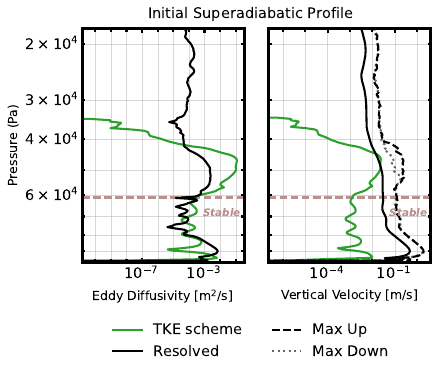}
    \caption{Eddy diffusivity (left panel) and vertical velocities (right panel) of the CM1 simulation initialized on a superadiabatic profile. In the left panel, the eddy diffusivity from the CM1 test case is calculated using Eqn.~\ref{eq:eddy-diffusivity} and shown by the black line. In the right panel, the RMS velocity, maximum upward and maximum downward velocity from the CM1 simulation is shown in black, sold, dashed and dotted lines respectively. The green line plots the eddy diffusivity and vertical velocity output from the CM1 sub-grid turbulence parameterization. The brown dashed lines marks the convectively stable region.}
    \label{fig:ad-superad-kzz}
\end{figure}

We ran a CM1 case initialized on a saturated, superadiabatic profile to test whether superadiabatic states where $q_{v,\mathrm{sat}} \ge q_{\rm crit}$ remain stable over time. The superadiabatic simulation was performed since the adiabatic and isothermal simulations discussed previously show long radiative relaxation timescales and did not converge to a superadiabatic temperature profile. 

The superadiabatic simulation shows no convective mixing or significant vertical velocities throughout the entire simulation. Fig.~\ref{fig:adiabatic_results}(c) shows the temperature profile is steepening in the full domain due to radiative cooling. In the superadiabatic simulation, the temperature profile remains superadiabatic throughout the entire simulation, but is cooling towards the gray gas radiative equilibrium profile. 

As in previous cases, moisture is transported from the upper to lower atmosphere through condensation. Radiative cooling in the upper atmosphere produces condensates that rain out and evaporate in lower grid cells, keeping the entire atmosphere near saturation. As in the moist adiabatic case (b), condensates do build up at the surface. Vertical velocities are minimal within the full domain and primarily driven by evaporation at the surface and gravity waves. 

Fig.~\ref{fig:ad-superad-fluxes} shows the sensible and latent heat fluxes for the superadiabatic CM1 simulation. The latent heat flux shows a large peak at the surface due to surface based evaporation and a smaller peak above the top of the convective inhibition layer where water vapor is being transported upwards as in the dry then moist adiabatic (a), and moist adiabatic (b) CM1 simulations. The sensible heat flux is zero in most of the domain and only peaks in correlation with the latent heat flux, indicating mixing is solely driven on small scales by condensation and evaporation in the superadiabatic simulation. 
 
Fig.~\ref{fig:ad-superad-kzz} shows the eddy diffusivity in the left panel and RMS vertical velocity in the right panel for the superadiabatic CM1 simulation. Both $K_{\rm zz}$ and the RMS velocity are 1–2 orders of magnitude lower in the stable region compared to the dry then moist adiabatic (a), and moist adiabatic (b) CM1 simulations. As in the previous cases, above the convective inhibition layer, the sub-grid TKE scheme (green line) $K_{\rm zz}$ is larger than the resolved (black line) $K_{\rm zz}$ suggesting upward mixing of water vapor at the boundary of the stable layer is driven by sub-grid scale turbulent mixing. Neither the $K_{\rm zz}$ or RMS velocity recover or strongly increase above the inhibition layer suggesting convection is suppressed in the full domain. Above the stable layer there is a slight increase in $K_{\rm zz}$ and RMS velocities, but generally, vertical velocities and turbulent mixing remain small throughout the domain. Near the surface a local minimum in $K_{\rm zz}$, coinciding with the lifted condensation level, could be diagnostic and does not, by itself, imply stable stratification.  Assuming the latent heat flux transports the heat in the atmosphere and a vertically constant latent-heat flux, $K_{\rm zz}$, should have a local minimum  \citep[e.g.][]{ge_heat-flux-limited_2024}. However, Fig.~\ref{fig:ad-superad-fluxes} shows a near-surface spike but overall suppression of latent and sensible heat fluxes, consistent with reduced turbulent mixing in this simulation.

\subsection{Radiative Timescales}
\label{sec:radiation}

\begin{table}[h!t]
    \caption{Radiative timescales for all CM1 simulations, estimated using (i) the diffusive gray gas approximation (Eqn.~\ref{eq:t-rad}) and (ii) a time-stepping solver that iteratively integrates the gray gas radiative transfer equations.}
    \label{table:greycond-timescale}
    \centering
    \begin{tabular}{lcc}
    \toprule
        & Diffusive & Time-stepping  \\
        & [yrs.] & [yrs.] \\ 
    \midrule
        \multicolumn{3}{l}{\textbf{Initial Isothermal Simulations }}  \\ 
    \midrule
        (b) Isothermal $q_v \approx q_{\rm crit}$ & 4   & 4  \\
        (c) Isothermal $q_v > q_{\rm crit}$       & 13  & 1  \\
        (d) Isothermal $q_v < q_{\rm crit}$       & 15  & 4  \\
    \midrule
        \multicolumn{3}{l}{\textbf{Initial Adiabatic Simulations}}    \\ 
    \midrule
        (a) Dry then Moist Adiabat & 4    & 1  \\
        (b) Moist Adiabat          & 5    & 5  \\
        (c) Superadiabatic         & 9    & 6  \\
    \bottomrule
    \end{tabular}
\end{table}

The radiative relaxation timescale for the CM1 simulations presented in this work is long, preventing the simulations from converging to a statistical equilibrium state. Across the CM1 simulations, the column-mean temperature tendency is $\mathcal{O}(10^{-5}\text{–}10^{-7})\ \mathrm{K\,s^{-1}}$. At these rates, a $1\,\mathrm{K}$ adjustment requires roughly $1.2$–$116$ days $\big(1/10^{-5}\approx 1.2\ \text{days},\ 1/10^{-7}\approx 116\ \text{days}\big)$, highlighting the long radiative relaxation timescales. We estimate the radiative timescale using (i) the diffusive gray gas approximation and (ii) a time-stepping solver that iteratively integrates the gray gas radiative transfer equations. 

The diffusive radiative timescale required to reach gray gas radiative equilibrium is  
\begin{equation}
    \label{eq:t-rad}
    t_{\text{rad}} \sim \frac{p} {g} \frac{c_p}{4  \sigma T^3},
\end{equation}
where $p$ is the pressure, $c_p$ is the specific heat capacity, $g$ is the gravity, $T$ is the temperature, and $\sigma$ is the Stefan-Boltzmann constant \citep{showman_atmospheric_2002}. We evaluate Eqn.~\ref{eq:t-rad} for both the isothermal and adiabatic CM1 simulations using the domain- and time-mean $p$ and $T$. Table~\ref{table:greycond-timescale} reports the maximum value from the profile. 

Additionally, we estimate the radiative timescale by time-stepping the final CM1 column towards a gray gas radiative equilibrium, defined by a height-independent net flux $F = I_{+}-I_{-}$ (i.e., $dF/dz=0$). The time-stepping calculation is performed in python and is initiated with the final state profiles from the CM1 simulations. Starting from the final CM1 temperature profile $T(p,0)$, we compute upward and downward two-stream fluxes and the radiative heating rate, $\partial T/\partial t$, from the gray gas Schwarzschild equations. We then update the temperature profile by, 
\begin{equation}
    T(p, t+1) = T(p, t) + \frac{dT}{dt} \cdot dt,
\end{equation}
where $T(p, t+1)$ is the new temperature profile, $\frac{dT}{dt}$ is the calculated temperature tendency, and $dt$ is the time step. The calculation holds the surface temperature fixed to the CM1 value (to mimic the fixed SST lower boundary condition in CM1) and keeps $q_v(p)$ and thus, $\tau(p)$ fixed in time. The new temperature profile is used to update $\partial T/\partial t$ using the gray gas radiative transfer equations. We iterate until the column maximum of $|\partial T/\partial t|$ is two orders of magnitude smaller than the final state CM1 heating rate. The elapsed model time provides our radiative relaxation estimate.

The time-stepping approach assumes that the atmosphere is purely radiative and neglects any further contribution from convection and dynamical feedbacks between convective and inhibited regions. Therefore, this calculation should be viewed as an indication of the time required for the inhibited regions to reach equilibrium. Table~\ref{table:greycond-timescale} reports the estimated radiative timescale from the two methods for all of the CM1 simulations. For each case, the final-state radiative fluxes and heating rates are shown in the Appendix.

%*******************************************************************************
\section{Conclusions} \label{sec:conclusions}

Understanding the atmospheric state of planets with $\rm H_2$-dominated atmospheres improves our ability to represent convection in global general circulation models and to better interpret observations of these worlds. While the simulations presented in this work focus on H$_2$O tracer in $\rm H_2$-dominated atmospheres, convective inhibition due to condensing compositional convection is relevant for any planetary atmosphere with a condensable tracer that has a higher mean molecular weight than the background atmosphere. Examples where condensation-driven convective inhibition is important include water and ammonia in sub-Neptune exoplanets, silicates in magma ocean planets \citep[e.g.][]{misener_importance_2022, markham_convective_2022}, methane and water vapor in Uranus and Neptune's atmosphere \citep{irwin_hazy_2022}, and water vapor in Saturn's atmosphere \citep{li_moist_2015}.

In this work, we use 3D, convection-resolving simulations to explore compositional-driven convective inhibition in H$_2$-dominated atmospheres. Any atmospheric tracer in a H$_2$-rich atmosphere is heavier than the background atmosphere. Therefore, in H$_2$-rich atmospheres, compositional gradients of atmospheric tracers can suppress convection, or inhibit convection if the tracer is condensing and the vapor mixing ratio exceeds a critical threshold, $q_{\rm crit}$ \citep{guillot_condensation_1995}. 

We test the condensation-driven convective inhibition 1D theory from \cite{guillot_condensation_1995} using two sets of CM1 simulations: (i) simulations initialized on saturated isothermal profiles and (ii) simulations initialized on adiabatic profiles. The isothermal runs exhibit rapid initial convection that adjusts the column toward a moist-adiabatic state where $q_v(p) > q_{\rm crit}(p)$, followed by slow radiative relaxation. The adiabatic CM1 simulations show the formation of a distinct layer where convection is inhibited, and $q_v(p) > q_{\rm crit}$. Returning to the questions posed in the introduction: 
\begin{itemize}
    \item \textbf{\textit{Is convection inhibited when the tracer abundance surpasses the critical threshold defined by \cite{guillot_condensation_1995}?}} Both the isothermal and CM1 simulations show convective inhibition in saturated layers where $q_v(p) > q_{\rm crit}$, as evidenced by near-zero vertical velocities, reduced eddy diffusivity $K_{zz}$, and the sharp decline in sensible heat fluxes in the inhibited region. All three adiabatic simulations show convection remains suppressed above the inhibition layer suggesting a transition region above the convective inhibition layer where velocities and transport of vapor remains small and suppressed.
    
    \item \textbf{\textit{If convective inhibition regions form, is the temperature profile in these regions superadiabatic?}} The temperature profile and temperature difference plots in Fig.~\ref{fig:adiabatic_results} suggest that the temperature profiles steepen toward radiative equilibrium within and above the inhibition layer. Generally, we find that moist convection in the upper atmosphere is driven by radiative cooling and the upper atmosphere convection weakens over time as moisture is transported to the deep atmosphere.
    
    \item \textbf{\textit{How are heat and moisture transport impacted if convection is inhibited?}} Both the isothermal and adiabatic simulations show net moisture transport, driven by condensation and rainout, from the upper to the lower atmosphere between the initial and final states. This process builds moisture in the deep atmosphere for the dry-then-moist adiabatic case, and at the surface for the other CM1 test cases. The CM1 simulations also suggest that latent heat flux and turbulent mixing do transport a small amount of water vapor and heat through convectively inhibited layer, but it is generally too weak to sustain strong moist convection in the upper atmosphere.
\end{itemize}

A caveat of this work is that all the simulations have relatively long radiative relaxation times, and the results have not converged to a statistical-equilibrium state. To confirm if saturated, superadiabatic temperature profiles are stable according to the Guillot criterion, we performed a CM1 simulation initialized on a saturated superadiabatic state and found it remains stable against convection throughout the simulation time. Generally, the radiation modeling setup used in this work is simplified, and we do not consider scattering, a cloud opacity or cloud-radiative feedbacks, all which should be considered in more detail in future work along with more realistic radiation schemes.

\cite{leconte_3d_2024} performed similar 3D convection-resolving simulation to determine if condensation-driven convective layers form in H$_2$-rich atmospheres and to characterize the resulting atmospheric state. The modeling framework and simulation setup in this work differ from that used by \cite{leconte_3d_2024}.  \cite{leconte_3d_2024} employed a correlated-k radiation scheme, did not use any sub-grid turbulence schemes, and implemented a condensation scheme where condensation occurs when $q_v(p) > q_{\rm v, sat}(p)$ and evaporation occurs at the boiling point. Conversely, in our study, we use a gray-gas radiation scheme, implemented a sub-grid turbulent mixing parameterization, and use a simpler condensation scheme based on \cite{rotunno_airsea_1987}. Both studies find that convection is inhibited in saturated layers where $q_v(p) > q_{crit}$ in H$_2$-dominated atmospheres, and that evaporation occurs at the boundary of the inhibited layer. However, in this work, we find weak water vapor transport to the upper atmosphere and weakening of the moist convective layer aloft, whereas \cite{leconte_3d_2024} report stronger upward vapor diffusion that sustains vigorous moist convection in the upper atmosphere.

Our results suggest that condensing tracers can be depleted in the upper atmosphere of planets with H$_2$-rich atmospheres, potentially preventing these species from being detected in sub-Neptune atmospheres. Moist convection in the upper atmosphere could be restored by a giant storm in which deep convection mixes the troposphere, as proposed by \cite{li_moist_2015}. After a giant storm, the tropospheric temperature profile would reset to an adiabatic state, and radiative cooling would then restart the process of cooling the atmosphere until the next storm occurs. Episodic convection in H$_2$-rich atmospheres has been suggested for the solar system giant planets \citep[e.g.][]{sugiyama_intermittent_2011,  sugiyama_numerical_2014, li_moist_2015, seeley_resolved_2025, clement_storms_2024} and should be followed up for sub-Neptunes atmospheres. 

The Guillot moist convection inhibition criterion is derived locally and assumes full saturation. Likewise, connection-resolving simulations predict local, small-scale behavior and mixing rather than global scale mixing or circulation. Large scale flow, such as synoptic eddies and Hadley or Walker type circulations, can create sub-saturation, in which case the Guillot criterion would no longer apply. Large scale flows are not captured in convection-in-a box simulations, including our CM1 simulations. It is an interesting question how the Guillot theory plays out in a situation when dynamic sub-saturation is maintained over large regions of the atmosphere. Further, it is still unclear if superadiabatic radiative layers driven by condensation-driven convective inhibition are a planetary-wide feature. The global impact of condensation-driven convective inhibition on the mean atmospheric state and tracer transport remains to be quantified with dedicated global simulations or at minimum convection-resolving simulations that include the effects of large-scale circulations. For instance, how moist convection and storms that form where $q\_v(p) < q_{\rm crit}(p)$ interact with adjacent inhibited layers (e.g., via rainfall and downdrafts) is an open question \citep[e.g.][]{guillot_giant_2022}.

Further, \citep{leconte_condensation-inhibited_2017} and \cite{friedson_inhibition_2017} suggested double diffusive convection should also be inhibited in the Guillot regime. We do not test double diffusive convection in the simulations presented in this work, but double diffusive convection could have an important effect on vapor transport through inhibited layers, and should be followed up in future work. 

%*******************************************************************************
 \section*{Acknowledgments}

N.Habib would like to thank Tristan Guillot, Steve Markham,  Maxence Lefèvre and Hamish Innes for their helpful discussions in developing this work. This research has received support from the European Research Council (ERC) under the European Union’s Horizon 2020 research and innovation program (Grant agreement No. 740963 to EXOCONDENSE), and from the Alfred P. Sloan Foundation under grant G202114194 to the AETHER project. The authors would like to acknowledge the use of the University of Oxford Advanced Research Computing (ARC) facility in carrying out this work. We thank the anonymous reviewer for suggestions that improved the final manuscript.

%*******************************************************************************
\appendix

\section{Radiative Fluxes and Temperature Tendency from the CM1 simulations}
\begin{figure}
    \centering 
    \includegraphics[width=\textwidth]{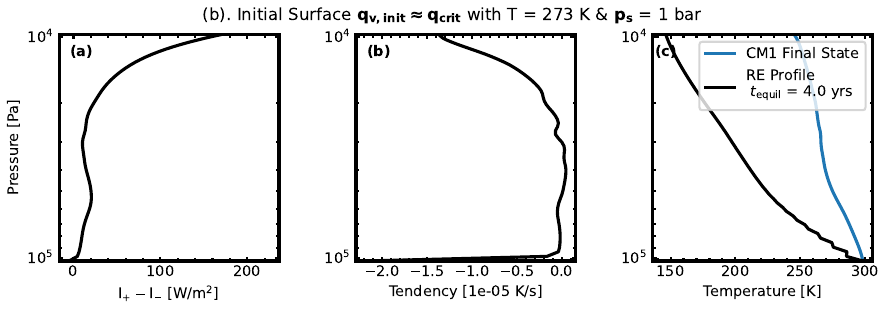}
    \includegraphics[width=\textwidth]{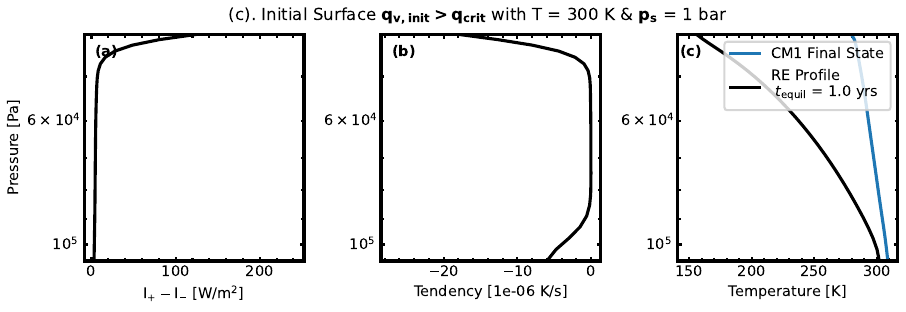}
    \includegraphics[width=\textwidth]{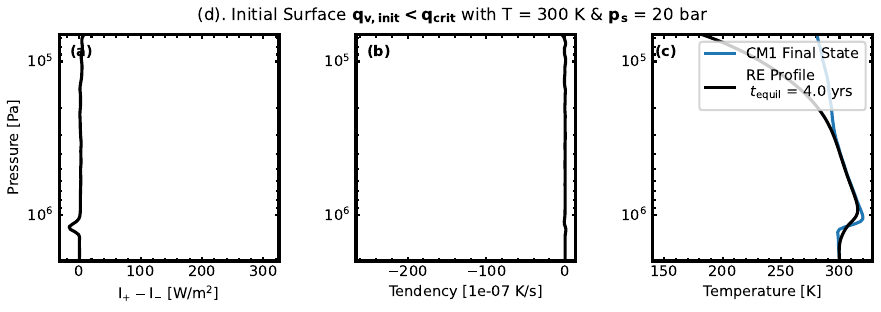}
    \caption{Radiative fluxes for the initially isothermal compositional convection CM1 simulations. Column (a) shows the difference of the upward and downward longwave infrared flux, $I_{+} - I_{-}$ and column (b) shows the temperature tendency, $\frac{dT}{dt}$, calculated by the TWOSTR numerical package in CM1. Column (c) plots the final state CM1 profile in blue and the temperature profile from the python time-stepping calculation which adjusts the final CM1 state towards radiative equilibrium (RE) in black.}
    \label{fig:isothermal-radiation-profiles}
\end{figure}

\begin{figure}
    \centering 
    \includegraphics[width=\textwidth]{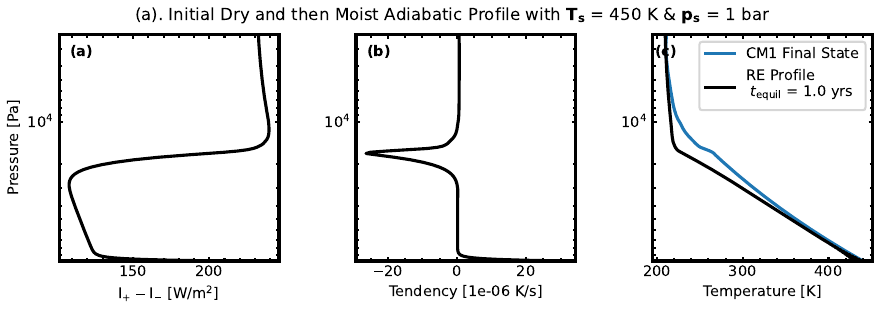}
    \includegraphics[width=\textwidth]{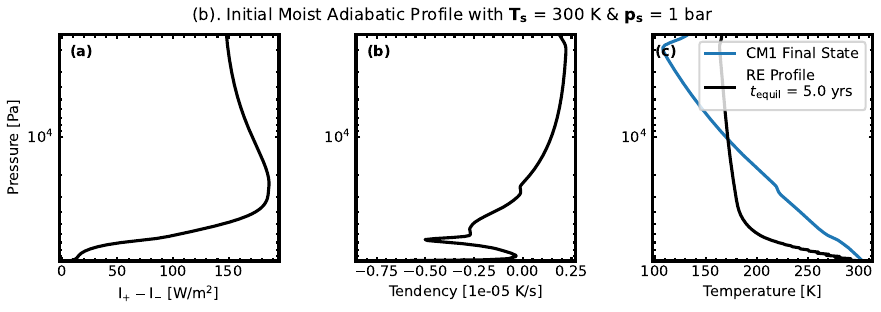}
    \includegraphics[width=\textwidth]{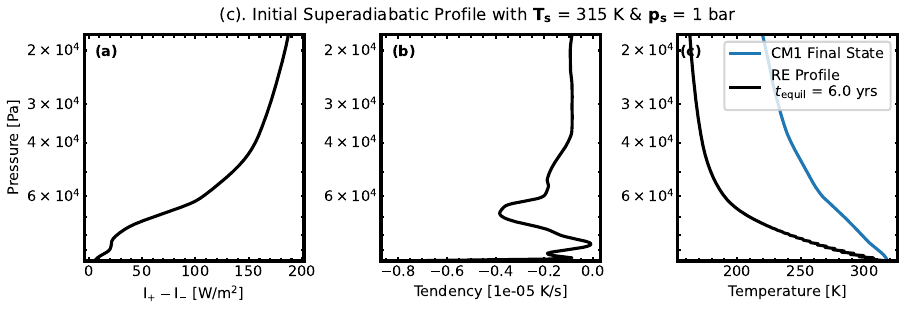}
    \caption{CM1 Radiative fluxes and estimated radiative timescale for initially adiabatic simulations. Column (a) shows the difference of the upward and downward longwave infrared flux, $I_{+} - I_{-}$ and column (b) shows the temperature tendency, $\frac{dT}{dt}$, calculated by the TWOSTR numerical package in CM1. Column (c) plots the final state CM1 profile in blue and the temperature profile from the python time-stepping calculation which adjusts the final CM1 state towards radiative equilibrium (RE) in black. The calculated radiative equilibrium profile provides an indication of the radiative relaxation timescale and the temperature for the layer where convection is inhibited.}
    \label{fig:adiabatic-radiation-profiles}
\end{figure}

We show an overview of the radiation fluxes and temperature tendency calculated by the TWOSTR numerical package in CM1 for the isothermal simulations in Fig.~\ref{fig:isothermal-radiation-profiles} and the adiabatic simulations in Fig.~\ref{fig:adiabatic-radiation-profiles}. In the left column, we plot the difference of the upward and downward longwave infrared flux, $I_{+} - I_{-}$ from the respective CM1 simulation, and in middle column, we show the temperature tendency, $\frac{dT}{dt}$ of the final state from the respective CM1 outputs. The temperature tendency is defined, 
\begin{equation}
    \frac{dT}{dt} = \frac{g}{c_{p, {\rm mix}}(p)} \frac{d(I_+ - I_{-})}{dp},
\end{equation}
where $p$ is the pressure, $T$ is the temperature, $t$ is the model time, $g$ is the gravitational acceleration, $c_{p, {\rm mix}}$ is the specific heat capacity of the mixture of background and tracer at a given pressure, and $I_+$ and $I_{-}$ are the upward and downward longwave infrared fluxes respectfully. We find in our CM1 simulations that the temperature tendency is on the order of $10^{-5}-10^{-7}$~K~s$^{-1}$, showing that the radiative relaxation timescale are long in all of our simulations. 

To quantify the radiative relaxation time, we performed a python time-stepping calculation where we estimate the time required to reach radiative equilibrium and to determine the corresponding temperature profile assuming convection is inhibited everywhere. In panel (c) of Fig.~\ref{fig:isothermal-radiation-profiles}~--~\ref{fig:adiabatic-radiation-profiles}, we plot the final temperature state from CM1 in blue, alongside the temperature profile that would result if the system were forced towards radiative equilibrium. The time-stepping calculation is described in Section~\ref{sec:radiation}. Fig.~\ref{fig:adiabatic-radiation-profiles} and Fig.~\ref{fig:isothermal-radiation-profiles} demonstrates that the CM1 simulation results have not yet achieved a state of statistical equilibrium, and that radiation still has the potential to modify the temperature profile, albeit over long timescales. 

\section{An Overview of the CM1 Model}

The equations solved within CM1 are presented and derived in detail in \cite{bannon_theoretical_2002} and \cite{bryan_benchmark_2002}, and are summarized within this Appendix. The CM1 code and a list of publications using this model are available at \url{https://www2.mmm.ucar.edu/people/bryan/cm1/}. CM1 has been used to model pure steam atmospheres \citep{tan_convection_2021} and cloud-convection feedback in the atmospheres of brown dwarfs \citep{lefevre_cloud-convection_2022}. The most extensive application of CM1 is the study of convective flows on Earth, covering topics such as tropical cyclones, thunderstorms, deep moist convection, and idealized studies aimed at enhancing our understanding of Earth's convection. 

CM1 is a non-hydrostatic fluid flow model that is used to simulate idealized atmospheric phenomena in a 3D Cartesian domain ($x, y, z$) as a function of time, $t$. The prognostic variables in CM1 are the 3D components of flow velocity ($u, v, w$), the non-dimensional pressure perturbation ($\pi'$), and the potential temperature perturbation ($\theta'$). The governing equations of CM1 are given by the ideal gas law, conservation of mass, energy, and momentum, per unit mass of the background atmosphere. 

The continuity equation for an atmospheric tracer with a mass mixing ratio $q_v$ in CM1 is given by, 
\begin{equation} \label{eq:conservation_tracer}
    \frac{Dq_v}{Dt} = -\dot{q}_{\rm cond} - \dot{q}_{\rm dep} - \frac{1}{\rho_{b}} \, \nabla \cdot (\rho_v \, v_{v}),
\end{equation}
where $\rho_b$ is the background atmosphere density, $\rho_v$ is the atmospheric tracer density, $v_v$ is the diffusive velocity of the atmospheric tracer (in vapor form, can be condensing or not) relative to the background atmosphere, $\dot{q}_{\rm cond} $ describes the rate at which vapor condenses to the liquid phase, $ \dot{q}_{\rm dep}$ is the rate of deposition, $\nabla$ is the del operator and $\frac{D}{Dt}$ is the material derivative. Generally, subscript $b$ refers to the background atmosphere while subscript $v$ refers to the atmospheric tracer in vapor form. The last term in Eqn.~\ref{eq:conservation_tracer} represents the diffusion of the atmospheric tracer in vapor form relative to the background air. The diffusion term is often small and is neglected in CM1 by default. Similarly, the continuity equation applied to the density of liquid, $\rho_l$, and solid, $\rho_s$, condensate components is given by, 
\begin{equation} \label{eq:conservation_tracer_liquid}
    \frac{Dq_l}{Dt} = \dot{q}_{\rm cond} - \dot{q}_{\rm freeze} - \frac{1}{\rho_{b}} \, \frac{\partial{(\rho_b V_l q_l)}}{\partial{z}},
\end{equation}
\begin{equation} \label{eq:conservation_tracer_solid}
    \frac{Dq_s}{Dt} = \dot{q}_{\rm dep} + \dot{q}_{\rm freeze} - \frac{1}{\rho_{b}} \, \frac{\partial{(\rho_b V_s q_s)}}{\partial{z}},
\end{equation}
where $z$ is the atmospheric height, $\dot{q}_{\rm freeze}$ describes the rate at which the liquid phase freezes, $q_l = \rho_l/\rho_b$ is the mass mixing ratio of the liquid condensate, $q_s = \rho_s/\rho_b$ is the mass mixing ratio of the solid condensate, and $V_s$ and $V_l$ are the terminal fall velocity of the solid and liquid condensates, respectively. The condensation, deposition, and freezing rates are determined by cloud microphysics schemes. 

Conservation of momentum is derived considering the gravitational force, the fluid pressure gradient, the Coriolis force, and frictional forces which we neglect in our study but are an optional parameter that can be added within CM1. Conservation of momentum, written in terms of $x\,, y\,, z$ components are, 
\begin{equation} \label{eq:conservation_momentum_x}
    {\frac{Du}{Dt} = -\frac{1}{\rho_{b}{(1 + q_{v} + q_{1} + q_{s})}} \, \frac{dp}{dx}} + f {v},
\end{equation}
\begin{equation} \label{eq:conservation_momentum_y}
    {\frac{Dv}{Dt} = -\frac{1}{\rho_{b}{(1 + q_{v} + q_{1} + q_{s})}} \, \frac{dp}{dy}} - f {u},
\end{equation}
\begin{equation} \label{eq:conservation_momentum_z}
    {\frac{Dw}{Dt} = -\frac{1}{\rho_{b}{(1 + q_{v} + q_{1} + q_{s})}} \, \frac{dp}{dz} + B };
\end{equation} 
where $ u, \; v, \, w$ are the $x\,, y\,, z$ components of velocity respectively, $f$ is the Coriolis parameter given by $ f = {2\Omega \sin \phi}$, $\Omega$ is the rotation rate of the planet, $p$ is the pressure, and $ B$ is the buoyancy of the air parcel. In this work, we neglect the Coriolis force and set $f$ = 0. The original CM1 code calculates the buoyancy as,
\begin{equation} \label{eq:buoyancy_dilute}
    B = g \left[ \frac{\theta'}{\theta_0} + (q_v - q_{v,0}) \left( \frac{1}{\epsilon} -1 \right) - (q_l - q_{l,0}) - ( q_s - q_{s,0}) \right].
\end{equation}
where $\theta$ is the potential temperature, $\epsilon$ is the mean molecular weight ratio of the tracer to the background air, and g is the gravitational acceleration. The subscript, $_0$ refers to a hydrostatic reference state while the $'$ denotes the perturbation from the base state at a give time. However, Eqn.~\ref{eq:buoyancy_dilute} assumes dilute amounts of $q_v$, $q_l$ and $q_s$. We have modified the CM1 code to use a more general form of buoyancy without making the dilute approximation, 
\begin{equation} \label{eq:buoyancy}
    B = - g \frac{\rho'}{\rho_0},
\end{equation}
where $\rho'$ is the total density perturbation at a given time and $\rho_0$ is the total density of the reference initial state. 

CM1 defines a non-dimensional pressure using a reference pressure, $p_{00}$, as
\begin{equation}\label{eq:pi}
    \pi \equiv \left( \frac{p}{p_{00}} \right) ^ \frac{R_{b}}{c_{p, b}},
\end{equation}
where $R_b$ is the specific gas constant of the background atmosphere and $c_{p, b}$ is the heat capacity at constant pressure per unit mass of the background atmosphere. Correspondingly, the potential temperature can be defined in terms of $\pi$ as,
\begin{equation}\label{eq:theta}
    \theta = \frac{T}{\pi}, 
\end{equation}
where $T$ is the temperature. The equation of state can be written in terms of $\pi$ as, 
\begin{equation}\label{eq:eos}
    \pi = \left( \frac{\rho_b R_b \theta (1 + q_v/\epsilon)}{p_{00}} \right) ^ \frac{R_{b}}{c_{v, b}},
\end{equation}
where $c_{v, b}$ is the heat capacity at constant volume per unit mass of the background atmosphere. 

Energy conservation is expressed by equations that govern the evolution of the non-dimensional pressure and potential temperature:
\begin{equation} \label{eq:theta_conservation}
\begin{aligned}
     \frac{D \theta}{Dt} = & - \left( \frac{R_{\text{mix}}}{c_{v,\text{mix}}} - \frac{R_{b} \, c_{p,\text{mix}}}{c_{p,b} \, c_{v,\text{mix}}}\right) \, \theta \nabla \cdot \bar{u} \\
     & + \left( \frac{c_{v,b}}{c_{v,\text{mix}} c_{p,b} \pi} \right)(L_v \dot{q}_{\text{cond}} + L_s \dot{q}_{\text{dep}} + L_f \dot{q}_{\text{freeze}}) \\ 
     & - \theta \frac{R_{v}}{c_{v,\text{mix}}} \left(1 - \frac{R_{b}{c_{p,\text{mix}}}}{c_{p, b} R_{\text{mix}}} \right)(\dot{q}_{\text{cond}} + \dot{q}_{\text{dep}}) + \dot{q}_{\text{rad}} + \dot{q}_{\text{dis}}
\end{aligned}
\end{equation}
where $R_v$ is the gas constant of the atmospheric tracer, $R_{\rm mix}$, ${c_{v,\text{mix}}}$, and ${c_{p,\text{mix}}}$ are the the gas constant, specific heat capacity at constant volume and pressure for a mixture of background air with an atmospheric tracer given by, 
\begin{equation}
    \rm x_{mix} = x_b \left( \frac{1}{1 + q_v} \right) + x_v \left(\frac{q_v}{1 + q_v}\right), 
\end{equation}
where $x$ denotes the thermodynamic property $R$, $c_p$, or $c_v$ and $x_{mix}$ can represent $R_{\rm mix}$, $c_{v, \rm mix}$, or $c_{p,\rm mix}$ depending on the property. Additionally, $L_v$, $L_s$, and $L_f$ denote the latent heat of vaporization, sublimation, and fusion respectively. $\dot{q}_{\text{rad}}$ is the potential temperature tendency due to radiation, and $\dot{q}_{\text{dis}}$ is the dissipative heating rate potential temperature tendency. The mass weighted mean velocity of the background and tracer atmospheric components, $\bar{u}$ is defined as, 
\begin{equation}
    \bar{u} = \frac{\rho_b u_b + \rho_v u_v}{\rho_b + \rho_v}, 
\end{equation}
where $u_b$ is the horizontal velocity of the background atmosphere, and $u_v$ is the horizontal velocity of the atmospheric tracer. 

In CM1 pressure is determined by,
\begin{equation} \label{eq:pressure_conservation}
\begin{aligned}
     \rm \frac{D \pi}{Dt} &= - \left( \frac{R_{b}}{c_{p, b}} \, \frac{c_{p,\rm mix}}{c_{v,\rm mix}}\right) \, \nabla \cdot \bar{u} \\
     & + \frac{R_{b}}{c_{p,b} c_{v,\rm mix} \theta} \, (L_v \dot{q}_{\text{cond}} + L_s \dot{q}_{\text{dep}} + L_f \dot{q}_{\text{freeze}}) \\  
     & - \frac{R_{b}}{c_{p,b}} \left(\pi \frac{R_{v}{c_{p, \rm mix}}}{R_{\text{mix}} c_{v, \text{mix}}} \right)(\dot{q}_{\text{cond}} + \dot{q}_{\text{dep}}) + \dot{q}_{\text{rad}} + \dot{q}_{\text{dis}}.
\end{aligned}  
\end{equation}
Eqns.~\ref{eq:pressure_conservation} and \ref{eq:theta_conservation} are used in CM1 to determine the thermodynamic state of the system.

CM1 allows for optional parameters which include tendencies of sub-grid turbulence, additional optional diffusive tendencies, Newtonian relaxation, cooling and warming due to condensate drag during fallout. CM1 integrates the governing equations using a third order Runge-Kutta time differentiation and fifth order spatial derivatives for the advection terms. The equations are evaluated in a two-step process. In the \emph{dynamical step} all terms involving time are integrated forward, neglecting any phase change terms. The dynamical step is followed by the \emph{microphysical step} where only terms involving phase changes are evaluated. CM1 then iterates between the two steps until $\theta$ converges. Once $\theta$ converges, CM1 moves on to the next time step. How the governing equations are formulated and integrated numerically is further discussed in \cite{bryan_benchmark_2002}, \cite{bryan_governing_2009}, \cite{bryan_sensitivity_2012}. The CM1 governing equations assume: the atmosphere behaves as an ideal gas, diffusive effects in the vapor phase are negligible, condensates are incompressible and have the same temperature as surrounding background atmosphere, and all heat capacity terms are independent of temperature.

\section{Dry Superadiabatic Simulation}

\begin{figure}
    \centering 
    \includegraphics[width=0.5\textwidth]{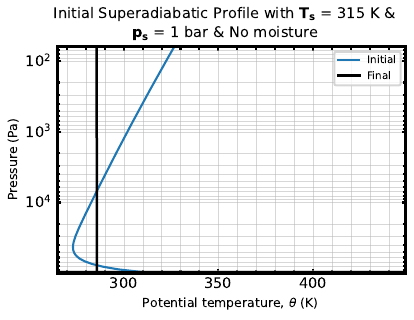}
    \caption{Potential temperature for a CM1 simulation initialized on a superadiabatic state but with no moisture with the initial state shown by the blue line and the final state shown by the black line. The dry superadiabat simulation was performed to benchmark the CM1 model performance.}
    \label{fig:dry-superad}
\end{figure}

We performed a CM1 benchmark case initialized with a dry, superadiabatic temperature profile to confirm the CM1 model behavior. The dry superadiabatic case uses the same setup as the moist superadiabatic simulation in the main text, but with moisture disabled (i.e., $q_v=0$). As shown in Fig.~\ref{fig:dry-superad}, the column adjusts to a dry adiabat, as expected. Without moisture there is no condensation to stabilize a superadiabatic profile, so convection mixes the layer to dry-neutral stratification.

\section{CM1 Simulation Lapse Rates}

\begin{figure}
    \centering 
    \includegraphics[width=1.0\textwidth]{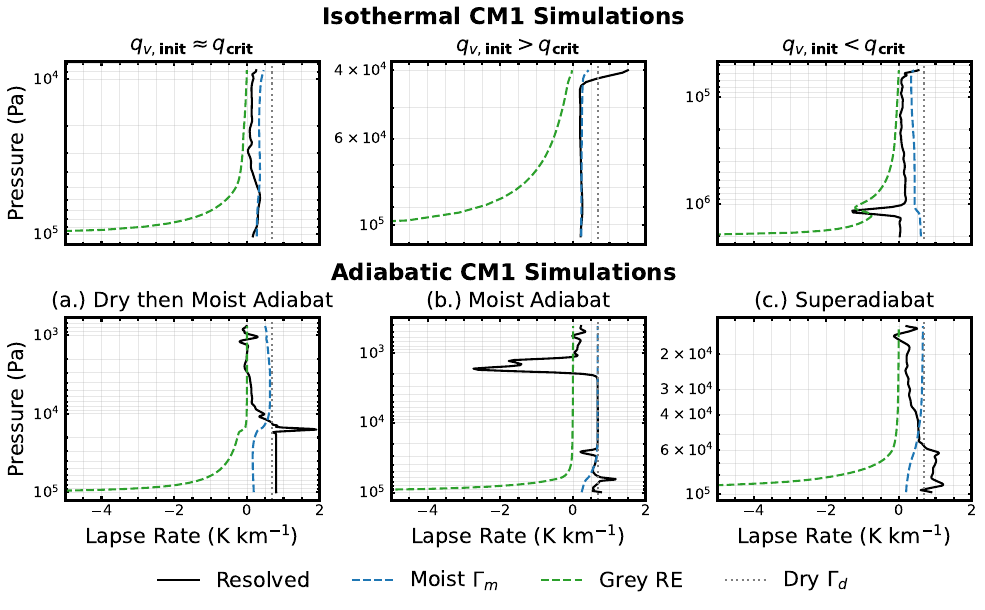}
    \caption{Lapse rates for all six CM1 simulations with radiation. In each panel, the dry-adiabatic lapse rate is shown by the gray dotted line, the moist-adiabatic lapse rate is given by the blue dashed line, and the gray-gas radiative-equilibrium lapse rate is shown by the green dashed line for comparison. The black line shows the lapse rate at the final timestep for the respective simulation.}
    \label{fig:lapse-rates}
\end{figure}

As a supplement to Figs.~\ref{fig:adiabatic_results} and \ref{fig:isothermal-results}, Figs.~\ref{fig:lapse-rates} shows the lapse rates for all six CM1 simulations with radiation. Each panel also includes reference curves for the dry-adiabatic lapse rate (gray dotted), the moist-adiabatic lapse rate (blue dashed), and the gray-gas radiative-equilibrium lapse rate (green dashed) for comparison.

\bibliography{references}
\bibliographystyle{aasjournal}

%*******************************************************************************
\end{document}